\documentclass[10pt,conference]{IEEEtran}
\IEEEoverridecommandlockouts

\usepackage{cite}
\usepackage{amsmath,amssymb,amsfonts}
\usepackage{algorithmic}
\usepackage{graphicx}
\usepackage{textcomp}
\usepackage{xcolor}

\usepackage{subcaption}

\begin{document}

\title{On the Future of Cloud Engineering}

\author{\IEEEauthorblockN{David Bermbach\IEEEauthorrefmark{1}, Abhishek Chandra\IEEEauthorrefmark{2}, Chandra Krintz\IEEEauthorrefmark{3}, Aniruddha Gokhale\IEEEauthorrefmark{4}, Aleksander Slominski\IEEEauthorrefmark{5},\\ Lauritz Thamsen\IEEEauthorrefmark{1}, Everton Cavalcante\IEEEauthorrefmark{6}, Tian Guo\IEEEauthorrefmark{7}, Ivona Brandic\IEEEauthorrefmark{8}, Rich Wolski\IEEEauthorrefmark{3}}
\\
\IEEEauthorblockA{\IEEEauthorrefmark{1}TU Berlin; Berlin, Germany; [david.bermbach,lauritz.thamsen]@tu-berlin.de}
\IEEEauthorblockA{\IEEEauthorrefmark{2}University of Minnesota; Twin Cities, MN, USA; chandra@umn.edu}
\IEEEauthorblockA{\IEEEauthorrefmark{3}UC Santa Barbara; Santa Barbara, CA, USA; [ckrintz,wolski]@ucsb.edu}
\IEEEauthorblockA{\IEEEauthorrefmark{4}Vanderbilt University; Nashville, TN, USA; a.gokhale@vanderbilt.edu}
\IEEEauthorblockA{\IEEEauthorrefmark{5}IBM T.J. Watson Research Center; Yorktown Heights, NY, USA; aslom@us.ibm.com}
\IEEEauthorblockA{\IEEEauthorrefmark{6}Federal University of Rio Grande do Norte;  Natal, Brazil; everton.cavalcante@ufrn.br}
\IEEEauthorblockA{\IEEEauthorrefmark{7}Worcester Polytechnic Institute; Worcester, MA, USA; tian@wpi.edu}
\IEEEauthorblockA{\IEEEauthorrefmark{8}Vienna University of Technology;  Vienna, Austria; ivona.brandic@tuwien.ac.at}
}

\maketitle

\begin{abstract}
Ever since the commercial offerings of the Cloud started appearing in 2006, the landscape of cloud computing has been undergoing remarkable changes with the emergence of many different types of service offerings, developer productivity enhancement tools, and new application classes as well as the manifestation of cloud functionality closer to the user at the edge. The notion of utility computing, however, has remained constant throughout its evolution, which means that cloud users always seek to save costs of leasing cloud resources while maximizing their use. On the other hand, cloud providers try to maximize their profits while assuring service-level objectives of the cloud-hosted applications and keeping operational costs low. All these outcomes require systematic and sound cloud engineering principles. The aim of this paper is to highlight the importance of cloud engineering, survey the landscape of best practices in cloud engineering and its evolution, discuss many of the existing cloud engineering advances, and identify both the inherent technical challenges and research opportunities for the future of cloud computing in general and cloud engineering in particular.
\end{abstract}

\begin{IEEEkeywords}
Cloud Engineering, Cloud Computing\footnote{This paper is the author copy of a paper published in the IEEE International Conference on Cloud Engineering (IC2E 2021); The text and figures are identical to the official version on IEEE Explore. \copyright  IEEE.}
\end{IEEEkeywords}

\section{Introduction}
Cloud Computing is a term coined for service-oriented computing on cluster-based distributed systems.
From a user perspective, it is an attractive utility-computing paradigm based on Service-Level Agreements (SLAs), which has experienced rapid uptake in the commercial sector.
Following the lead of Amazon Web Services (AWS)\footnote{aws.amazon.com}, many Information Technology vendors have since developed ``utility,''``cloud,'' or ``elastic'' product and/or service offerings -- from IaaS to SaaS~\cite{paper_lenk_cloud_landscape} or even human work as micro tasks~\cite{bermbach2011extendable,paolacci2014inside}.
Apart from specific feature set differences, all cloud computing infrastructures share two common characteristics: they rely on operating system virtualization (e.g., Xen, VMWare, etc.) for functionality and/or performance isolation and they support per-user or per-application customization via a service interface, which is typically implemented using high-level language technologies, APIs, and Web services.

This highly customizable, service-oriented methodology offers many attractive features.
Foremost, it simplifies the use of large-scale distributed systems through transparent and adaptive resource management, and simplification and automation of configuration and deployment strategies for entire systems and applications.
In addition, cloud computing enables arbitrary users to employ potentially vast numbers of multicore cluster resources that are not necessarily owned, managed, or controlled by the users themselves.
By reducing the barrier to entry on the use of such distributed systems, cloud technologies encourage creativity and implementation of applications and systems by a broad and diverse developer base.

Today, cloud engineering, i.e., the process of building systems and applications for cloud environments, can be considered a relatively mature field.
In fact, being ``cloud-based'' can safely be assumed to be the default for the majority of newly implemented applications.
Also, there have been years of research devoted to cloud engineering, e.g., published in the IEEE International Conference on Cloud Engineering (IC2E).
The computing world, however, is constantly changing.
New technologies such as Docker and Kubernetes, which revolutionized application packaging and deployment, have been developed. Emerging application domains such as IoT and AI have imposed new requirements on the cloud beyond those for traditional Web applications. Cloud engineers today have access to a vast range of higher-level cloud services. Clouds have become increasingly more decentralized with geo-distributed cloud infrastructures; this trend being further exacerbated by the emergence of edge computing.
All this means that cloud engineering as a discipline has undergone major changes and continues to do so.

In this position paper, we -- the organizers and the steering committee members of IC2E -- identify significant trends that we expect to dramatically change, or which we have observed already changing cloud engineering.
We start by giving an overview of today's best practices in cloud engineering (Section~\ref{sec:bestpractices}).
Then we discuss the main trends which we have identified and their implications for cloud engineering (Section~\ref{sec:trends}) before concluding the paper.

\section{Best practices in cloud engineering\label{sec:bestpractices}}

Cloud engineering introduces a number of new challenges in a software engineering, operations, and maintenance context.  Cloud applications typically invoke existing network-facing services through published API and amalgamate the resulting functionality, possibly also serving it through an API.  As a result, APIs form the units of computational and storage composition which raises the level of programming abstraction, and places reliability, performance, and maintenance requirements on the services that export them~\cite{paper_bermbach_benchmarking_web_apis,paper_bermbach_webapibenchmarking2}.

Thus, several new practices have emerged for engineering cloud applications and services that take these new requirements into account. These practices principally address four important characteristics of cloud-hosted services and applications: lifecycle, scale, composition, and cost.

Cloud services are often long-lived.  At the same time, because the software does not package and ship to a distributed community of customers, it can be updated ``on-the-fly’’ and transparently to its users.  Moreover, users interact with services via APIs.  As long as the APIs are stable (or accretive) in terms of their functionality, users are unaware (and indeed cannot become aware) of the service implementations.  These features allow cloud applications and services to be far more responsive to changing user requirements than shipped and packaged software.  

As a result, \textbf{Agile Software Engineering}~\cite{beck2001manifesto} processes have become the predominant software engineering methodology.  An Agile development process is one in which small to medium-sized teams constantly evolve the software to meet an ever-changing set of requirements.  Rather than engage in a complete requirements gathering and then a software specification process ahead of all development and testing, Agile development gathers requirements constantly and incrementally changes specifications, so that the software is never ``too far’’ out of step with the latest set of user needs and/or expectations.  Developers work with specifications (called tasks) that are scoped so that they can be completed over short time durations (typically two weeks or less).  A set of tasks, once completed, forms a user story which describes a specific user experience that the software is designed to support.  Ideally, user stories can be completed during a single (short) development cycle.  As a result, the software is always ready to ``release’’ partial functionality (described by user stories) at the end of each development period.
Agile software engineering embodies the notion that the lifecycle of the cloud service is typically far longer than the lifetime of functionality associated with any specific user requirements.  That is, the software is never ``done,’’ but rather always in a constant state of modification reflecting the changing needs of users and applications.  

Technologies such as \textbf{Continuous Integration/Continuous Deployment} (CI/CD) ``pipelines’’ have emerged to support the more responsive software engineering processes that cloud engineering fosters.  A CI/CD pipeline is a set of services that developers use to make updates to a shared codebase.  Rather than ``locking’’ the code base, CI/CD systems rely on intelligent merging operations so that concurrent updates can take place continuously and any conflict can be identified immediately.  The design of these merging operations is not well-understood in a general sense.  Ideally, all merges are managed by the CI/CD pipeline automatically, but, in practice, the development and testing teams must devote considerable effort to designing merge operations and practices for each software project.

Another feature of CI/CD technologies is that each software merge triggers one or more delivery operations.  Since Agile promotes ``ready-to-ship’’ development, the goal of most CI/CD pipelines is to test the ``readiness to ship’’ with each merge event and to provide immediate feedback to developers as to the status of their latest code contributions to the codebase.  Designing the altering and feedback mechanisms for a specific CI/CD instantiation is also not a well-understood general process.  If developers use the CI/CD pipeline as intended, they can merge contributions rapidly and frequently.  At the same time, such merges will not necessarily leave the software in a ``ready-to-ship’’ state.  Accurately capturing the state of the software from merge to merge is currently a bespoke practice for most software projects. Another challenge is how to assert quality of service of a group of software artifacts~\cite{grambow_continuous_benchmarking_2019,van2012kieker,paper_grambow_benchmarking_microservices,grambow_benchmarking_2020}.

\begin{figure}
    \centering
    \includegraphics[width=\columnwidth]{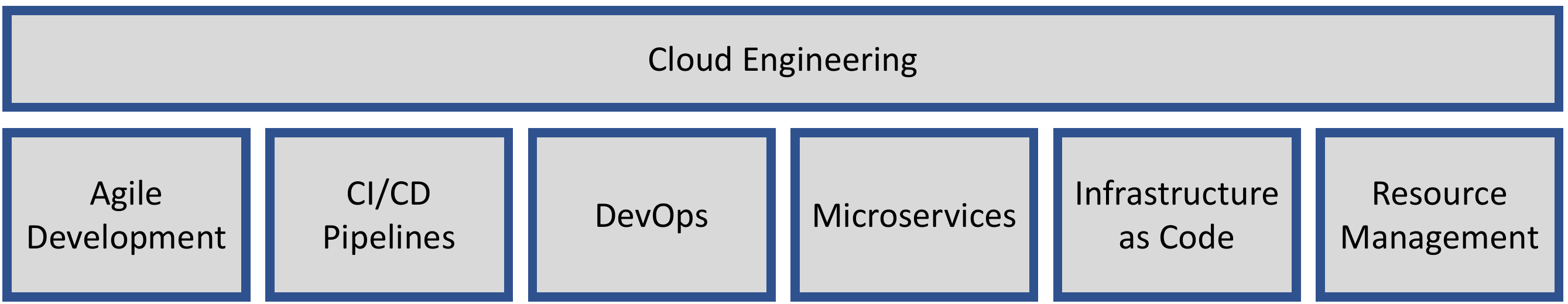}
    \caption{Main pillars of cloud engineering.}
    \label{fig:buildingblocks}
\end{figure}

\textbf{Resource Management} and scale (in terms of resource count, i.e., scaling in/out as well as up/down) is also an aspect of cloud engineering that has required new technological approaches.  More properly, it is not the scale of the resources that can be incorporated into a cloud service or application but rather the \textit{speed} with which the scale can be changed up or down.  In particular, clouds automate the provisioning and de-provisioning of resources and services under programmatic control.  Because humans need not be ``in the loop’’ when resources are committed or released, cloud engineering must consider how, when, and why resources are automatically provisioned and/or de-provisioned.  Thus, cloud developers must often consider developing automation control as part of their development activities~\cite{roy2011efficient,shekhar2018performance}.  Further, because scaling responses may be driven by user activity (and not physics), successful feedback-control methodologies from other disciplines (e.g., chemistry, aeronautics, signal process, etc.) are often ineffective or failure-prone.  Further, unlike the services themselves, testing this control system, at some level, can require a significant commitment of resources.  Future research that enables this development to take place economically in conjunction with the development of service functionality is needed.

Combining Agile development processes, CI/CD technologies to support those processes, and automated resource control has led to new maintenance practices for cloud engineered services and applications.  In particular, because developers must create code that controls the operation of the services when deployed and because CI/CD allows for frequent releases and service updates, many engineering organizations use a \textbf{DevOps} model~\cite{bass2015devops}.  In a DevOps setting, the development team includes members with operational skills (ideally, all developers on such a team have such skills) and the responsibility for running the software for its users is shared by the development staff rather than by a separate Information Technology (IT) organization.  

Conjoining development and operational responsibilities within the same team both fits the Agile and CI/CD process models more congruently (compared to a ``traditional’’ siloed development and IT organization) and improves the quality of testing during development (since developers are ultimately responsible for their code's operational stability).  The design and composition of DevOps teams, however, varies considerably from project to project.  Ideally, all developers share operational responsibilities equally.  However, from a practical perspective, developer and operational skill sets vary, often substantially, within a team.  The organizational principles that lead to effective DevOps teams, particularly when geographically distributed and at scale, as well as the technologies necessary to support such teams are both active areas of research.  

With service APIs (and not programming language primitives) as the fundamental unit of application program composition, code reuse becomes a challenge for cloud-engineered applications.  Specifically, APIs are high-level and, thus, often specialized to specific service functions; so, composing an application from multiple, previously implemented APIs can lead to sometimes unresolvable functional conflicts (between service dependencies) within an application.  

One approach to addressing this composition problem is to structure services so that their APIs are as ``narrow’’ and simple as possible.  These \textbf{Microservices} promote service reuse and effective testing since individual APIs necessarily implement simple functionality, but it creates challenges for the applications that are composing the APIs.  Specifically, the proliferation of APIs creates the need for developers to discover available microservices and determine the specific functionality associated with each API.  Because they are services (and not library calls), they are often stateful.  Thus, an application that calls a microservice, encounters the current ``state’’ of the service (i.e., the service does not necessarily restart from a known state when the application begins using it).  Reasoning about the current state associated with a proliferation of services within the application code creates significant debugging challenges. Resolving these challenges in a CI/CD context remains an important research question.

Finally, the rental model that most clouds implement as a charging policy for cloud usage requires that developers consider monetary cost as part of development.  Fixed capacity (non-cloud) data center resources operate on an amortization basis with respect to accounting.  The resources are charged for when they are provisioned and the ``up-front’’ cost is amortized (often including depreciation) over time.  Cloud resources, famously, are available on a ``pay-as-you-go’’ basis, meaning that as applications automatically provision resources, their owners are only charged for the resources that they provision.  Similarly, when an application releases resources, the recurring rental charges for those resources also end.  

The advantage of this approach is that, in principle, it is possible for an application to optimize the cost of the infrastructure it consumes well beyond what is possible in a fixed-capacity/amortization model.  The disadvantage is that developers must now reason about application infrastructure costs that fluctuate as the application uses cloud automation.  Furthermore, the adoption of microservices exacerbates this problem since each service may carry its own unique rental cost structure.  Simply \textit{predicting} what the monetary cost associated with an application will be is a significant research challenge.  Finally, large-scale clouds often use eventually consistent storage to implement their account reporting features.   Thus, even when a development team has engineered a cost monitoring and predicting control system for its application, it may need to rely on information that is many hours old.  If the automated resource provisioning features malfunction, building a purely reactive system to detect and stop significant cost overruns may not be possible. Thus, designing and implementing cost control features for cloud-engineered services and applications remains an important research challenge that has not yet been fully addressed.

A side effect of the frequent provisioning and de-provisioning of resources -- both driven by the rental model and CI/CD pipelines -- is that cloud systems are deployed and torn down frequently. This makes manual system deployment prohibitively expensive and implies that the usual imperative installation scripts will frequently encounter errors which they are ill-equipped to cope with. As a solution, so-called \textbf{Infrastructure as Code} (IaC) has emerged. In IaC, developers specify the desired deployment outcome (resources, installed systems) in a declarative way, while the IaC framework asserts that the desired state is reached~\cite{kuroda2014model,bhattacharjee2018model}. Chef, Puppet, Terraform, and Ansible are examples of widely used IaC frameworks. See Fig.~\ref{fig:buildingblocks} for a high-level overview of the main pillars of today’s cloud engineering.

\section{Ongoing and Future Trends in Cloud Engineering\label{sec:trends}}
In this section, we will give an overview of what we perceive as the main trends in clouds and how they affect cloud engineering.

\subsection{Decentralizing the world}
Traditionally, the cloud is about economies of scale, thus leading to centralization.
This means that today a large part of the world's applications and data resides in the data centers of a few major players.
While this has obvious benefits, it also has a number of disadvantages, which is why initiatives will approach this from various angles or have already started to do so.
For one, government agencies in charge of antitrust laws and regulations will be increasingly interested in the activities of the leading cloud players.
On the other hand, blockchains as a grassroots movement towards decentralizing computation have evolved.
In either case, the implication is that future cloud users will often work with multiple cloud providers -- this is usually referred to as cloud federation~\cite{paper_kurze_cloud_federation}.
Typically, cloud federation software abstracts the specifics of particular cloud services, providing its own APIs and formats to use cloud services and specify required service levels.
This addresses vendor lock-in, as users develop their applications less against a specific cloud provider’s environment, but instead against a federation library, having the federation software translate and encapsulate the specifics of multiple cloud platforms.
Moreover, many cloud federation tools also automatically select and combine cloud services from multiple providers, given specifications of required services and service levels, automating allocation tasks and raising the level of abstraction.
For example, cloud federation tools can provide access to virtual machines, storage, networking, as well as higher-level cloud services and APIs, taking into account prices, SLAs, and data protection regulations~\cite{Salama_2014_QoSFederation,Rebai_2015_Federation,Hiller_2018_CPPLFederation}.
The cloud federation approach has recently received considerable attention in Europe, especially around the GAIA-X project\footnote{https://www.gaia-x.eu/} and as a counter proposal to the hypercentralization into very large data centers and very few cloud providers.
In this context, we expect important research to be conducted and novel approaches to be developed that help users make sense of and optimally utilize multiple cloud offerings from edge to cloud infrastructures and different providers.

\subsection{Edge and fog: the cloud is no longer isolated}
The cloud offers many highly useful properties such as elastic scalability, on-demand billing, low total cost, and the illusion of infinite resources.
At the same time, however, the cloud is often quite far from the end-users, which means that end-users will often experience relatively high latency when interfacing with the cloud.
While this is acceptable for the Web-based workloads native to the cloud, it introduces problems for emerging application domains such as interconnected driving, e-health, or even smart homes~\cite{paper_bermbach_fog_vision}.
Another problem besides latency is bandwidth limitations: even today, it is completely impossible to send all created data to the cloud for processing.
The majority of data are, hence, discarded~\cite{paper_zhang_gdp,paper_pfandzelter_functions_streams}.
Furthermore, this problem is bound to become more pronounced since the number of sensors and IoT devices is growing faster than network bandwidth.
Finally, centralized clouds are problematic from a privacy perspective:
When all data are stored in the same location in the cloud, linking information from different sources can be done fast -- even at scale~\cite{paper_pallas_fog4privacy}.
\begin{figure}
    \centering
    \includegraphics[width=\columnwidth]{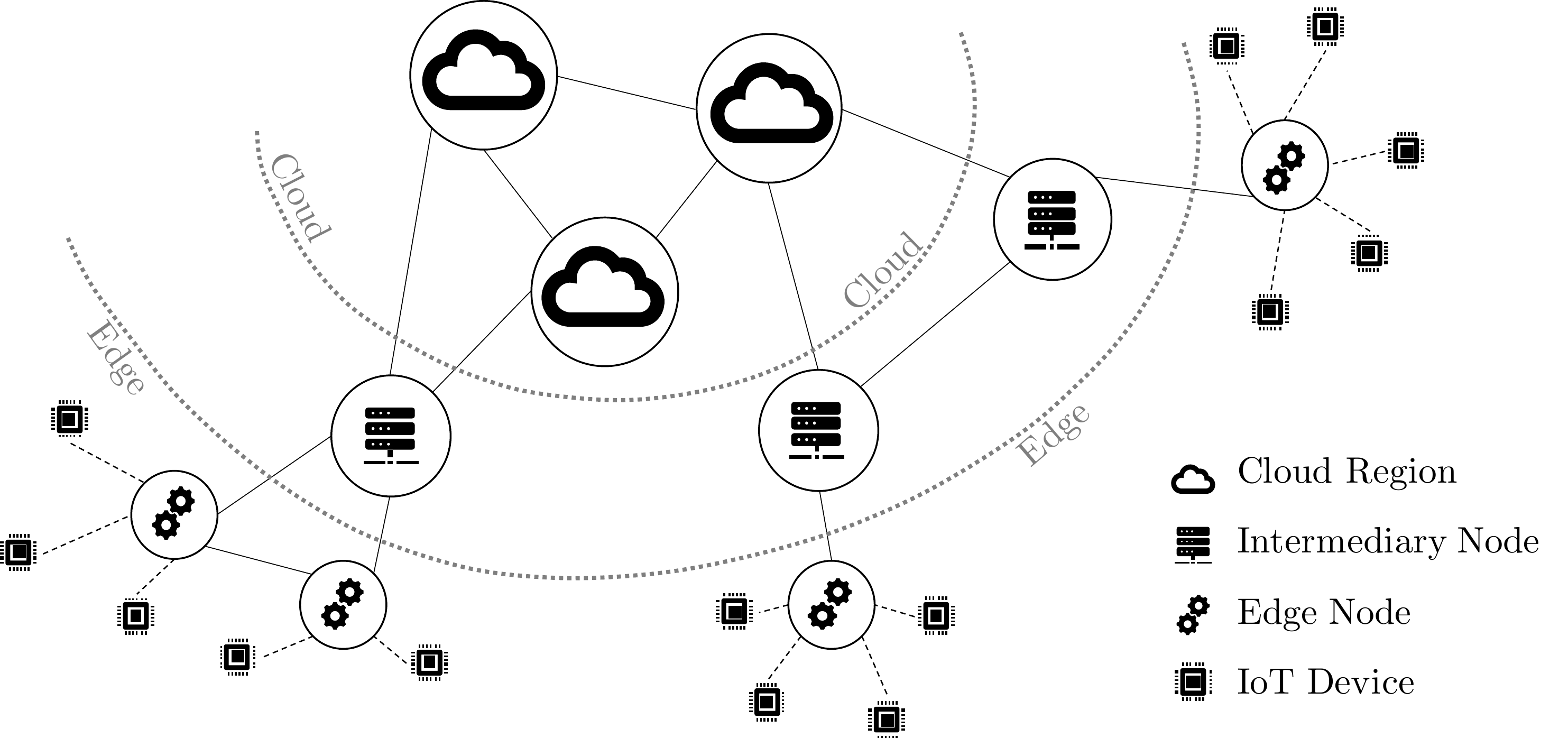}
    \caption{A layered architecture of cloud, edge, intermediary fog nodes, and IoT devices (Figure source:~\cite{paper_pfandzelter_zero2fog}).}
    \label{fig:fogarch}
\end{figure}

All these problems have at least partially been addressed with the emergence of fog and edge computing in which additional compute nodes closer to end-users (``edge nodes'') as well as on the way towards the cloud (see Figure~\ref{fig:fogarch}) are combined with existing cloud services~\cite{paper_bermbach_fog_vision}:
Latency is addressed by having parts of applications and data closer to end-users, bandwidth limitations are addressed by preprocessing and filtering data at the edge~\cite{shekhar2017indices,shekhar2019urmila}, and privacy is supported through strategies such as decoupled data hubs or multi-staged filtering~\cite{paper_pallas_fog4privacy}.

For cloud engineering, this means that the cloud is no longer the only place where application code is running and that runtime environments are becoming much more heterogeneous. In this sense, parts of the application and its data may be deployed on the edge, in the network between edge and cloud, or even on embedded devices~\cite{paper_george_nanolambda,qiu2019monocular}.
This means that cloud systems need to be designed for various runtime environments, need to be ready for migration without much pre-warning, need to be able to disable resource-hungry features when running on more constrained nodes, and need to be able to tolerate much more frequent failures.
For data management, this is particularly challenging due to the wide area replication~\cite{paper_zhang_gdp,paper_hasenburg_towards_fbase,techreport_hasenburg_2019,confais2017object}.
For this, we can likely reuse past research on cloud federation, e.g.,~\cite{paper_kurze_cloud_federation,paper_bermbach_cloudfederation}.

\subsection{LEO Internet: bringing data directly into the cloud}
\begin{figure}
    \centering
    \includegraphics[width=\columnwidth]{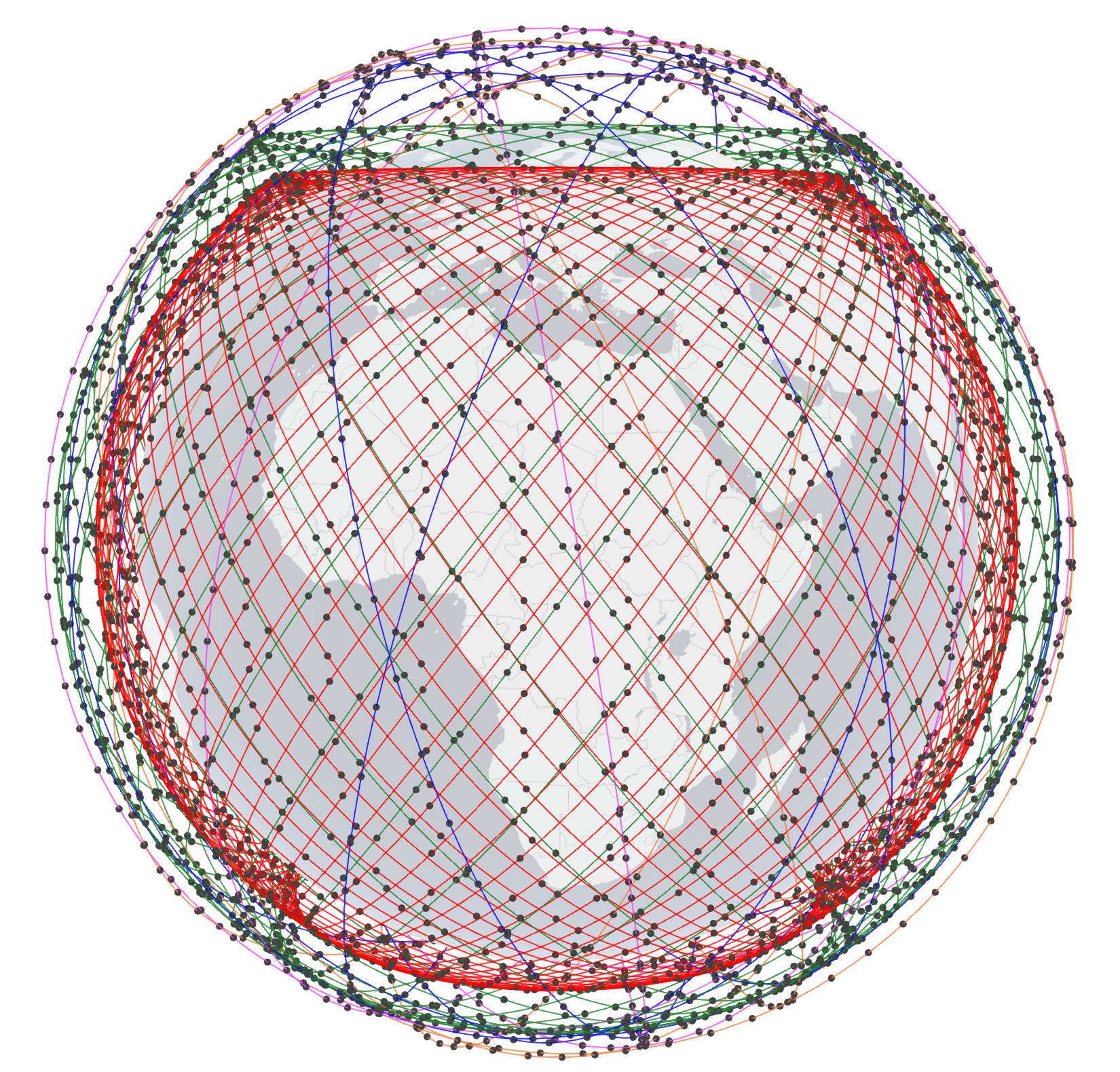}
    \caption{Overview of Starlink's phase I LEO satellite constellation (figure source:~\cite{paper_pfandzelter_LEO_serverless}).}
    \label{fig:constellation}
\end{figure}

\begin{figure}
    \centering
    \captionsetup[subfigure]{justification=centering}
    \begin{subfigure}{.5\columnwidth}
        \centering
        \includegraphics[height=3.5cm]{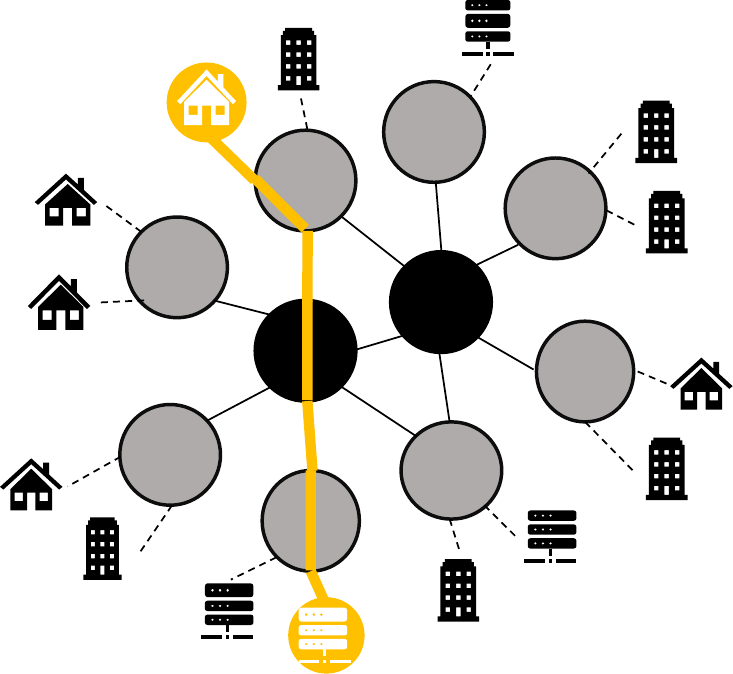}
        \caption{Traditional network topology.}
        \label{fig:topology_hierarchical}
    \end{subfigure}%
    \begin{subfigure}{.5\columnwidth}
        \centering
        \includegraphics[height=3.5cm]{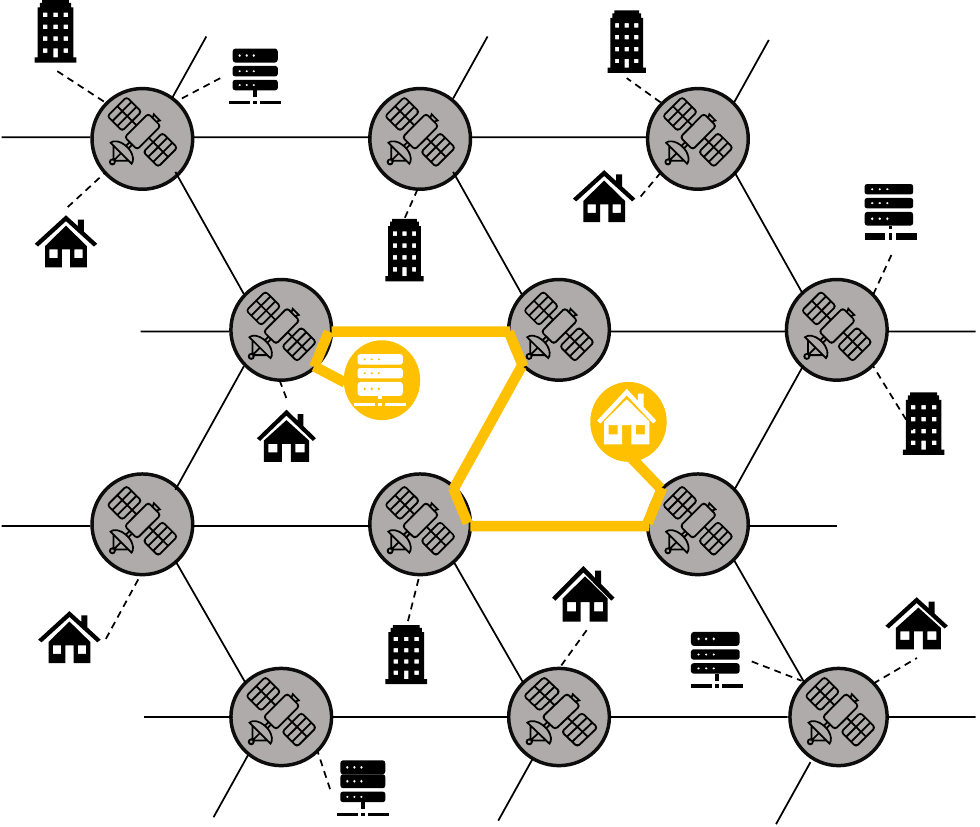}
        \caption{LEO Internet topology.}
        \label{fig:topology_sat}
    \end{subfigure}%
    \caption{In traditional networks, requests traverse a hierarchical tiered topology. In the LEO Internet, clients communicate directly via ground-stations and inter-satellite links which form a dynamic grid-like topology (figure source:~\cite{paper_pfandzelter_LEO_CDN}).}
    \label{fig:topology}
\end{figure}

Traditionally, the connection from end-users and devices to the cloud is via cables (and possibly via radio for the last mile).
This also allows network providers to insert edge and fog intermediary nodes on the path from end device to cloud.
On an abstract level, the existing networks resemble a set of trees that are interconnected near their root nodes via the Internet backbone.
Today, we see an alternative way emerging in which low Earth orbit (LEO) satellites interconnected in a grid layout can directly be accessed anywhere on Earth~\cite{Pultarova2015-ml} (see Figure~\ref{fig:topology}).
Such LEO constellations, e.g., the ones deployed by SpaceX' Starlink (see Figure~\ref{fig:constellation}) or Amazon's Project Kuiper, leverage their low orbits to provide low latency, high bandwidth Internet access.
From a cloud provider perspective, this is highly interesting because they can directly connect end devices to the cloud while they are ``yet another leaf'' in the tree-based fiber Internet.
In fact, this might be one of the reasons behind Amazon's Project Kuiper.

The first generation of LEO satellites are essentially dumb pipes: a ground station connects to a satellite, satellites forward the requests via inter-satellite links until it is downlinked to the destination ground station.
It is, however, very likely that the next generation of LEO satellites will include compute infrastructure~\cite{otte2013f6com,levendovszky2013distributed,balasubramanian2015drems,Bhattacherjee2020_kr,Bhosale2020_aa}, e.g., for running application code closer to end users~\cite{Bhosale2020_aa,paper_pfandzelter_LEO_serverless} or to reduce bandwidth usage via a LEO-based content delivery network~\cite{paper_pfandzelter_LEO_CDN}.

For cloud engineering, this means that existing systems do not only need to be extended towards edge and fog (or work together with dedicated systems), but that they also have to be extended to the LEO edge, i.e., a LEO satellite with compute capabilities.
Key challenges there will be very limited resources (e.g., due to space restrictions) and having no physical access for servicing the device.
Finally, LEO satellites are permanently moving, i.e., are not geostationary.
Many systems rely on always connecting clients to the same server, e.g., for caching or sticky sessions.
Hence, creating virtual geostationarity is arguably the main challenge.

\subsection{IoT and AI are becoming the main applications}
When the cloud started to become popular, it was primarily used for Web-based applications, which could thus cope with being slashdotted by quickly scaling their application resources. The second application type that emerged in the cloud was what was later called big data, e.g., the New York Times' TimesMachine~\cite{web_timesmachine}, as large amounts of resources could be provisioned and released on-demand. In the third wave, we saw businesses move their traditional enterprise systems to the cloud to save costs. While all these applications still run in the cloud, we nowadays see many IoT and AI applications appear in the cloud.
In the following, we will briefly describe five types of such applications -- both for cloud-only and mixed cloud/edge environments.

\subsubsection{Data collection at the edge}

Currently, we are experiencing a paradigm shift towards the so-called tactile Internet applications driven by the emergence of single-digit millisecond latency in mobile 5G networks.
In such applications, we usually use IoT devices (e.g., sensors) that are rather tiny and unable to run complex computation.
Processing and data storage is handled on more powerful edge nodes in the vicinity of the IoT devices that are capable of running machine learning (ML) models locally on the edge, e.g., as in edge ML~\cite{cit-edgeMLGoogle}.
After local decisions based on the IoT data, data are often aggregated and moved to the cloud for further processing~\cite{paper_pallas_fog4privacy,paper_pfandzelter_zero2fog}, e.g., to retrain ML models or for secondary use of such data.

The main challenge for cloud engineering is that IoT devices are often low-cost hardware with high failure rates.
Furthermore, sensor aging, software failures, or various network problems may lead to missing or incorrect data.
In the short term, such data failures will lead to wrong decisions on the edge. When data, however, are propagated to the cloud, such flawed data will affect decisions in the long term.
Hence, data forwarded from the edge to the cloud should always be taken with a grain of salt and should never be assumed to be complete, correct, and up to date.
While there are first approaches in this regard, e.g.,~\cite{8941259} which uses ARIMA and exponential smoothing, dealing with data quality issues remains a key challenge for cloud engineering.
Another challenge, in the case of IoT applications which often rely on pub/sub, is the question of data transport: Where are brokers deployed, where are messages filtered, how do brokers interact, e.g.,~\cite{paper_hasenburg_disgb,paper_hasenburg_geobroker,paper_hasenburg_broadcast_groups}.

\subsubsection{Geo-distributed Data Analytics}

As more and more data is generated by end-users and IoT devices, there is an increasing need for
analyzing this data to extract useful, timely information. However, much of this data
is generated at the edge and is highly geo-distributed. Collecting and aggregating all
the data to a centralized data center is infeasible due to bandwidth, cost, and latency constraints.
As a result, there have been research efforts towards building efficient geo-distributed data analytics systems and algorithms~\cite{pu_low_2015,heintz16mr,vulimiri2015wanalytics,viswanathanAA16,hung_wide_area_2018}.
Cost is an important consideration for cloud users; there is a wide diversity in the cost of computing, storage,
and networking costs within and across cloud providers, that must be taken into account~\cite{kimchi}. 
 A key research challenge is to dynamically identify the right combination of cloud providers, data centers, and edge resources for data analytics to 
provide the user-desired cost-performance tradeoff.

For continuously-generated sensor and IoT data, timely analysis is essential to generate actionable results.
This requires streaming analytics systems that are designed for geo-distributed environments, with many dispersed data sources as well as
distributed computation resources.
Such streaming analytics systems must be able to utilize both edge resources (for timely in-situ processing) and 
cloud resources (for aggregating distributed data). Research efforts have focused on designing stream computing
systems for geo-distributed environments~\cite{jetstream2014,heintz2015optimizing,zhang_awstream_2018} that can identify the best strategies to
place and schedule analytics tasks across multiple data centers. Recent work has developed 
mechanisms for adaptability and fault tolerance~\cite{jonathan20wasp}, multi-query optimizations~\cite{albert_multi_query} that can take advantage
of common data and operators, and algorithms to achieve the best latency-traffic-accuracy tradeoff~\cite{heintz_trading_2016,heintz_optimizing_2017,ttl_aggregation} in generating
timely results. A key research challenge is to extend such efforts to
highly heterogeneous edge-cloud environments and to support diverse applications from 
traditional query processing to video/image processing and machine learning.

\subsubsection{Model training and distribution from cloud to edge}
In a geo-distributed setting, e.g., as in edge computing, ML models are usually trained in the cloud before distributing a smaller or reduced version of the model to the edge for inference in the vicinity of end-users and devices~\cite{bhattacharjee2019stratum}. When deploying ML models over geographically distributed edge nodes, non-stationarity arises as a challenging problem. Due to environmental changes, models that have been learned and trained and finally distributed to edge nodes might become inaccurate and, in the worst case, not valid anymore. Inefficient model (re-)distribution might become a performance bottleneck.
In traditional data centers, non-stationarity is solved using so-called online learning, where models are trained in batches as new data arrives. Applying online learning in a geo-distributed setting bears several problems in terms of sustainability, where distributed ML models can be independently trained and periodically synchronized through a centralized parameter server. The frequency of such synchronization controls a tradeoff between model staleness and network load.
While there are some approaches for this, e.g.,~\cite{DBLP:conf/hotos/CiparHKLGGKX13,DBLP:journals/pomacs/AralEB20}, a key challenge for cloud engineering is the question of when and where to train and update models and how to distribute them to the edge~\cite{bhattacharjee2020deep}.

\begin{figure}
    \centering
    \includegraphics[width=\columnwidth]{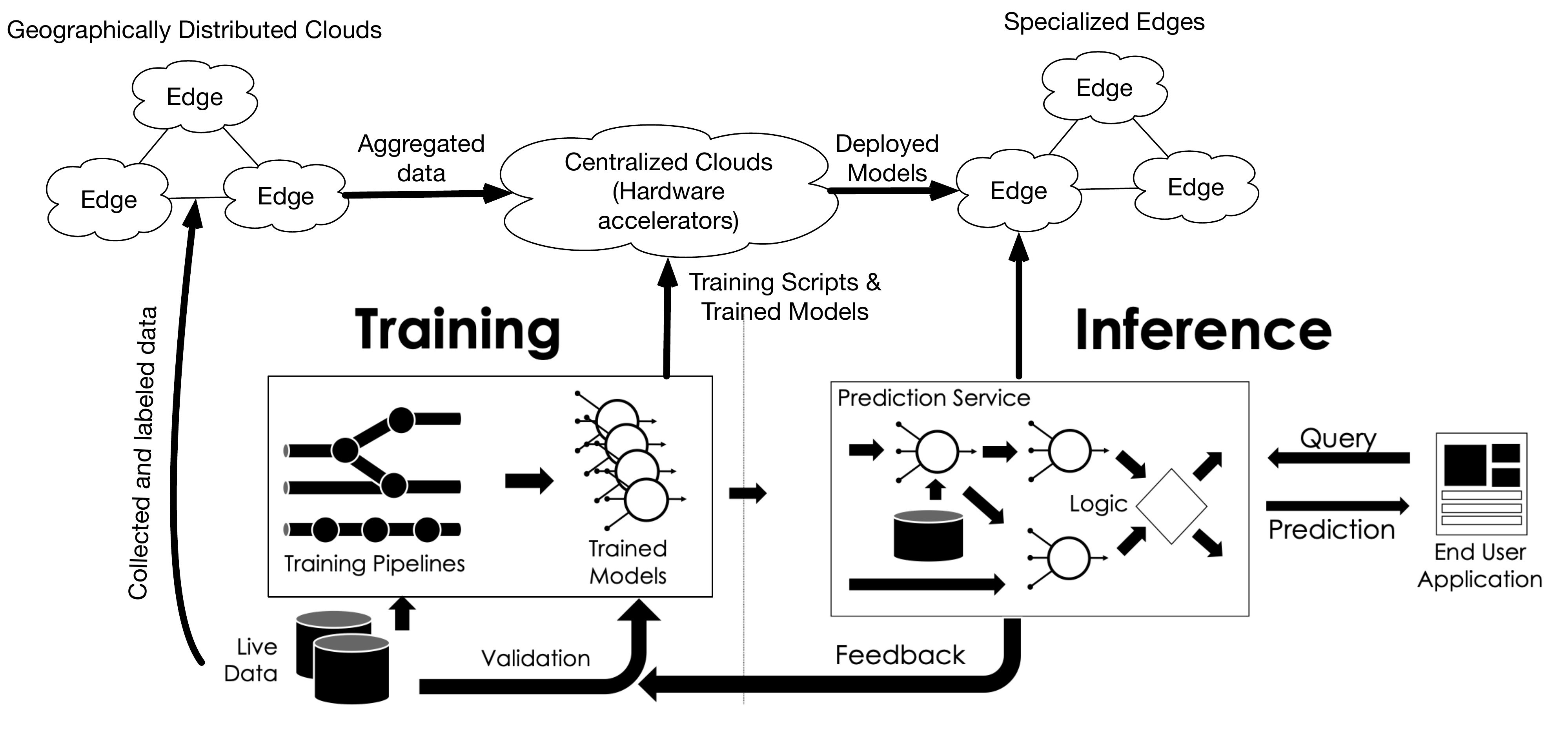}
    \caption{Clouds play an integral role in supporting ML lifecycles from data collection, model training, and model storage to inference serving (figure is adapted from~\cite{UCB_ML_slides}).}
    \label{fig:cloud_dl}
\end{figure}

\subsubsection{Training deep learning models in the cloud}

As cloud providers started to provide GPU access, it enabled deep learning practitioners to train larger deep neural networks (DNNs) which are otherwise difficult to train with limited cluster GPU resources.
To leverage distributed resources for training larger DNNs with terabytes of data, practitioners often need to resort to either \textit{model parallel} or \textit{data parallel} model learning approaches.
The former is often used when model memory requirements surpass the single-GPU memory, while the latter allows distributing processing of mini-batches to different GPU nodes.
As such, one of the key questions is how to synchronize model parameters among different nodes for high training throughput and converged accuracy.
Prior work innovated in the SGD protocol design space and investigated questions such as when to send gradients~\cite{zhao2019dynamic,dutta2020slow,li2021syncswitch_icdcs} and what gradients to send~\cite{Aji2017-yr,Wen2017-uf,Alistarh2017-ss}.
With the need for training DNNs beyond convolutional neural networks~\cite{Thorpe2021-vh,Wang2019-uy}, DNN training remains a challenging cloud engineering problem -- even with the emergence of serverless training services~\cite{Carreira2019-kx,Jiang2021-wa}.
Additionally, as the training scenario shifts from dedicated clusters (one training job per cluster) to shared clusters, problems such as resource provisioning (with cheap transient resources)~\cite{Li2020-bk,Harlap2017-wa} and GPU scheduling~\cite{Peng2018-mw,Gu2019-hc,Qiao2021-nk} still remain unresolved to effectively trade-off cluster utilization and training accuracy and throughput.
Edge resources and micro-data centers close to data sources can be utilized for distributed DL training~\cite{dlion}, which requires solving challenges of data distribution and resource heterogeneity.

\subsubsection{Deep learning inference in the cloud}
As deep learning models are widely deployed for user-facing cloud applications, we are witnessing a surge of inference workload.
We define \emph{inference workload} as executing one or many deep learning models to produce results for end-users~\cite{crankshaw2020-ut,Hauswald2015a, Capes2017a,google:translate:pipeline,bhattacharjee2019barista}.
Regardless of whether we are using CPUs or GPUs for inference executing, an inference request often goes through logical steps of loading models and waiting in queues.
As prior studies demonstrate, model loading time can be orders of magnitude higher than other inference time components using current deep learning frameworks like TensorFlow or with serverless computing~\cite{Gilman2019-lw,Romero2021-sz,Fuerst2021-bz,bhattacharjee2019barista,zhou2020cost}.
One na\"{\i}ve solution to reduce model loading time is to keep models in memory; this might work well for popular models but can lead to low resource utilization for less popular models.
The resource utilization problem is further exacerbated in the cloud when needing to manage many deep learning models of different popularity in parallel.
An interesting research question is then \emph{how to manage a large number of deep learning models given dynamic workload to satisfy performance objectives while lowering resource costs~\cite{barve2019fecbench}}.
One promising direction is to design deep learning-specific caching algorithms to manage main memory resources and minimize the performance impact of the model cold start problem~\cite{Gilman2019-lw}.
Formulating the model management problem as a caching problem allows leveraging rich literature on caching; however, questions such as how to incorporate the relatively slow PCIe transfer to GPUs and effectively use virtual GPU memory remain unsolved.
Other potential directions for addressing model loading performance include devising partition schemes for unified memory and dynamic resource provisioning.

The accuracy of deployed deep learning models might gradually deteriorate due to reasons such as \emph{concept shifting}~\cite{Lu2020-qy}.
To maintain desired accuracy level, recent work often employs online training or offline retraining, sometimes called continual learning~\cite{Diethe2019-ag}.  
Despite the algorithm advancement for continuous learning, key questions such as how to monitor and detect accuracy degradation, when to schedule continuous learning jobs, and how to balance the resource requirements of training and inference jobs in shared clusters remained unsolved.
One challenge in detecting accuracy degradation is the lack of ground truth data. Recently, researchers have leveraged offline powerful deep learning models to exploit the ``benefit of hindsight'' to reduce the efforts in data labeling~\cite{cvpr2021:WAD:talks}.
Such approaches promise to prioritize improving model performance on more difficult corner cases by detecting errors after the fact and supplying new labeled data.
As real-time detection of accuracy degradation is challenging, prior work often resorts to heuristic approaches (e.g., periodically or driven by new training data) in determining when to trigger retraining~\cite{De_Lange2019-gp}.
Given that the benefit of continuous training is hard to quantify, it can be interesting to treat it as a best-effort workload and devise scheduling algorithms to manage shared clusters with a mix of interactive and best-effort workload.
While the question of retraining in mixed cloud/edge cases is mostly driven by the tradeoff between staleness and network load, the focus in cloud-only scenarios lies in detecting the need to retrain.

\subsection{Going serverless: FaaSification of the cloud}
\begin{figure}
    \centering
    \includegraphics[width=0.8\columnwidth]{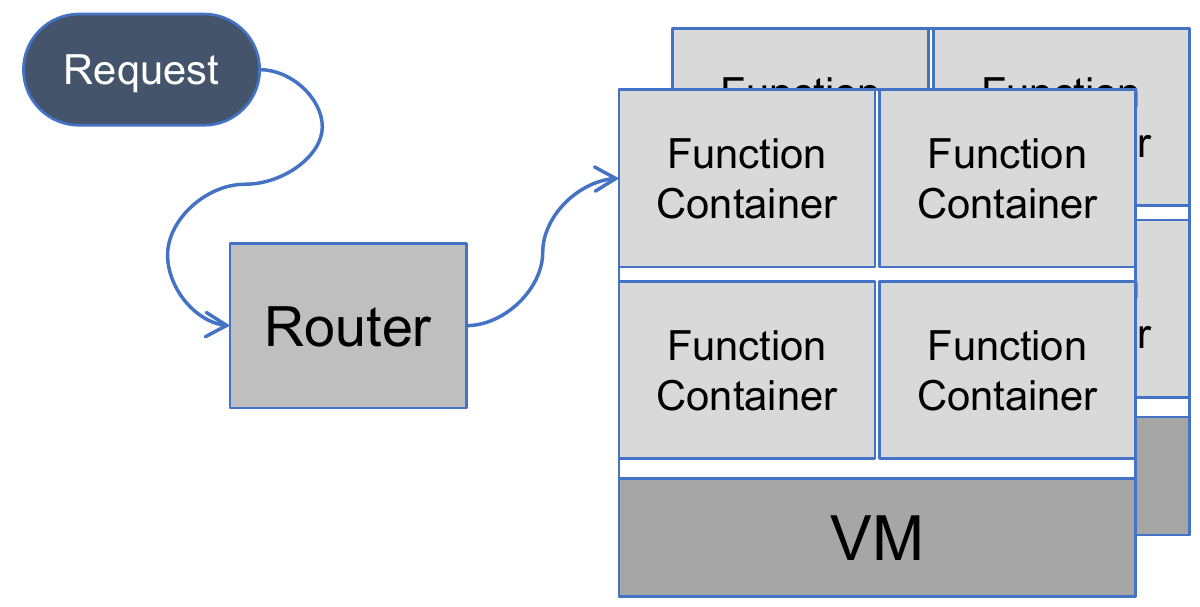}
    \caption{High-level overview of typical FaaS platforms: Requests are sent to a router component responsible for load balancing and authorization before being sent to a function container (either an actual Linux container~\cite{paper_baldini_openwhisk} or a microVM~\cite{paper_agache_firecracker_vms}) for execution on a VM or physical machine.}
    \label{fig:faasarch}
\end{figure}

Function-as-a-Service (FaaS, also known as cloud functions), is an emerging paradigm for cloud software development and deployment in which software engineers express arbitrary computations as simple functions that are automatically invoked by a cloud platform in response to cloud events (e.g., HTTP requests, performance or availability changes in the infrastructure, data storage and production, log activity, etc.) -- see also Figure~\ref{fig:faasarch} for a high-level architecture overview.
FaaS is the main building block of serverless computing, which is hence often used as a synonym for FaaS but combines FaaS with additional cloud services.
FaaS platforms automatically set up and tear down function execution environments on-demand (typically using Linux containers~\cite{paper_baldini_openwhisk,paper_pfandzelter_tinyfaas} or sometimes micro VMs~\cite{paper_agache_firecracker_vms}), precluding the need for developers explicitly to provision and manage servers and configure software stacks.
Developers construct and upload functions and specify triggering events.
Functions are typically written in high-level languages including Python, Java, or Node.js, leverage cloud services for their implementation, and communicate via HTTP or similar protocols.

FaaS applications are characterized by large numbers of transient, short-lived, concurrent functions.
Because the cloud (and not the developer) provisions the necessary resources, and such functions (by definition) can tolerate a high degree of multi-tenancy, application owners pay a very small fee (after any ``free tier'' usage) for CPU, memory, and cloud service use (e.g., \$0.20 per 1M invocations per month, and \$0.00001667 per memory * execution time).
To facilitate scale at a low price point relative to virtual server rental, cloud providers restrict function size (i.e., memory, code size, disk) and execution duration (e.g., 5 minutes maximum).

Amazon Web Services (AWS) released the first commercially viable FaaS, called AWS Lambda, in 2014~\cite{aws-lambda,aws-lambda-pricing}.
Since that time, the model has received widespread adoption because of its simplicity, low cost, scalability, and fine-grained resource control versus traditional cloud services.
Its popularity has spawned similar offerings in other public clouds (e.g., Google and Azure Functions) and open source settings, e.g., knix.io (previously known as SAND~\cite{paper_akkus_sand}), OpenWhisk~\cite{paper_baldini_openwhisk}, OpenLambda~\cite{paper_hendrickson_openlambda}, or the Serverless Framework~\cite{serverless_framework}.
Today, FaaS is used to implement a wide range of scalable, event-driven, distributed cloud applications, including websites and Cloud APIs, Big Data analytics, microservices, image and video processing, log analysis, data synchronization and backup, and real-time stream processing.

The FaaS programming paradigm simplifies parallel and concurrent programming. This is a significant step toward enabling efficiency and scale for the next-generation (post-Moore's-Law era) of advanced applications, such as those that interact with data and the physical world (e.g., the Internet of Things (IoT))~\cite{nsf-cps15,serverless-iot17}.
However, the complexity of asynchronous programming that these new applications embody requires tools that developers can use to reason about, debug, and optimize their applications.
Today, some simple logging services are available from serverless platforms to aid debugging.

FaaS opens up several cloud engineering questions regarding the operation and the use of FaaS platforms.
Operators of FaaS platforms have to deal with the cold start problem~\cite{paper_manner_coldstarts,paper_puripunpinyo_java,paper_lloyd_microservice,paper_lloyd_serverless}, which occurs when a request does not meet an idle function container/VM, and corresponding challenges in predicting request arrival.
Further challenges lie in function scheduling, especially under consideration of data and possibly with locally stateful functions~\cite{paper_hellerstein_serverless,paper_sreekanti_cloudburst}, and in function placement when the FaaS platform spans cloud and edge~\cite{paper_george_nanolambda,paper_bermbach_auctions4function_placement,paper_baresi_serverless_fog}.
We also expect developments that will close the gap between FaaS and streaming systems~\cite{paper_jain_splitserve,paper_pfandzelter_functions_streams}.
From the developer side, the main challenges are how to size functions or how to build applications using various composition approaches~\cite{paper_bermbach_faas_coldstarts,paper_ristov_afcl}, especially cross-provider, or how to benchmark FaaS platforms~\cite{paper_scheuner_faas_benchmarking,paper_grambow_befaas}.

\subsection{Ease of Cloud use: simplicity of serverless computing beyond FaaS}
Today, there is an abundance of cloud services and open-source tools available for developers.
This, however, does not make their life easier. It rather increases complexity since having so many options creates cognitive load (sometimes called choice overload or overchoice), which may lead to analysis paralysis: how can developers know that the choice they made is optimal? At the same time, developers are asked to deliver more functionality faster. This can only be achieved when using cloud services that are easy to use for developers. The majority of developers are not cloud engineering experts (and do not have time to become cloud engineering experts). It is estimated that there are about 27 million developers and only “4 million developers use cloud-based development environments”\cite{web_developers_worldwide} -- that means that the majority of professional developers (almost 75\%) are not cloud engineering experts.

From the beginning, the cloud promised on-demand access to computing resources~\cite{armbrust2010-vc} but did not address how to manage them when they are not used (the so-called “scale to zero”) or how to easily scale up and down in response to demand. The engineering tools were eventually provided, but developers were left with low-level building blocks -- for example, Netflix developed an internal tool to deal with AWS auto-scaling groups called Asgard that was eventually deprecated by Spinnaker~\cite{web_asgard_spinnaker} and complemented by Titus container management~\cite{web_titus} also developed by Netflix. Only big companies may afford to invest in building custom tools used by their developers to fully take advantage of cloud services.

There is a clear need to make the consumption of cloud services simpler for developers which is arguably the main driving force behind serverless computing gaining popularity.
As described above in the FaaSification section, developers, when using FaaS, only need to write functionality as functions that are invoked when needed and they are charged only for time when function code is running with (auto)scaling taken care of by cloud providers. The desire for simplicity is not limited to the business domain as similar complexities exist in the scientific domain~\cite{OccupyCloud2017}.

Serverless computing goes beyond FaaS and the same approach (pay-only-when-functionality-used and auto-scaling) is gaining popularity for other cloud services: cloud vendors are now marketing many of their services as serverless in areas such as Compute, Storage, Integration, Monitoring, Workflows, Devops, etc.
Examples include AWS Serverless\footnote{aws.amazon.com/serverless}, Azure Serverless\footnote{azure.microsoft.com/solutions/serverless}, or Google Cloud Serverless\footnote{cloud.google.com/serverless}.
Essentially, serverless computing makes building cloud applications as easy as building with LEGO: prefabricated bricks are assembled and connected with small custom code parts (FaaS functions) and most of the operational aspects (low-level cloud engineering) is left to the cloud providers.

For cloud engineering research, the key challenge is focusing on ease of use of cloud computing (and hence serverless computing). The ultimate goal is to allow developers who do not have cloud engineering expertise to get started and be productive in building cloud-native applications without becoming cloud experts. Only this way, the long-term promise of the cloud becoming like other utilities, such as the electrical grid, will be realized.

\subsection{Machine Learning plays an increasing role for cloud systems}

Given the scale of today’s cloud infrastructures, the large number of cloud services and application components that cooperatively respond to the requests of thousands of users, as well as the massive amount of concurrent tasks in parallel cloud jobs, cloud systems cannot be efficiently managed by human operators without appropriate tools. Therefore, increasing automation of management and operation tasks is required~\cite{bhattacharjee2019stratum}. For this, research and practice are increasingly turning to ML, training and applying new models for cloud resource management and cloud operation. Significant problems that have been addressed in this way include capacity planning, dynamic scaling and load balancing, scheduling and placement, log analysis, anomaly detection, and threat analysis.

A key area of work in this context focuses on having resource managers automatically adapt resource allocation, job scheduling, and task placement to the specifics of workloads, computing infrastructures, and user requirements. The goal is to reserve an adequate amount and type of resources for the required performance of jobs and have resource managers adjust to workload characteristics continuously by re-scaling resource allocation and scheduling jobs based on their resource demands onto shared cloud resources. Several approaches use performance models to provision and dynamically scale resources for data processing jobs~\cite{Venkataraman_Ernest_2016,Alipourfard_CherryPick_2017,Thamsen_Ellis_2017}, using either profiling runs or historic executions of recurring jobs to train scale-out models. Many other works apply reinforcement learning to integrate the exploration of potential solution spaces directly with an optimization towards given objectives such as high resource utilization, low interference, and cluster throughput. In this way, several novel cluster schedulers use either classical or deep reinforcement learning methods to schedule various types of cluster jobs in large data center infrastructures~\cite{Hongzi_2016_DeepRM,Cheong_2019_SCARL,Thamsen_2020_MaryHugoHugo}. Other systems use reinforcement learning, for example, to re-provision and scale microservices towards given service-level objectives~\cite{Qiu_2020_FIRM}. Another possibility is to apply techniques commonly utilized with recommender systems, such as collaborative filtering, for scheduling and placement in large cloud infrastructures~\cite{Delimitrou_Paragon_2013,Delimitrou_Quasar_2014}. While there is no consensus yet as to which methods work best for the different possible objectives and workloads, there is a clear trend towards using ML and increasingly also deep learning to optimize all aspects of resource management in data centers.

The second area of active research and development focuses on continuous monitoring, log analysis, and anomaly detection. Due to the scale of infrastructures and systems, human operators increasingly have difficulties to work with the sheer amount of monitoring data and logs generated in today's data centers. Therefore, a major trend is using ML to support cloud operations, also referred to as AIOPs, in which extensive monitoring is combined with stream processing and ML methods to automate operational tasks based on the state of systems. A central task is noticing any performance degradations and failures early on. Many specific examples of works in this area identify anomalies using time-series forecasting, online clustering, and other unsupervised methods on monitoring data, traces, and logs~\cite{Gulenko_2016_MLforClouds,Huang_2017_TSA,Ibidunmoye_2018_AAD,Nedelkoski_2020_Logsy}. Other approaches use, for instance, graph neural networks to identify and locate issues in connected microservices~\cite{Wu_2020_MicroRCA,Scheinert_2020_Telesto}.
A closely related task is to automatically remediate issues and threats once they have been identified and before they lead to severe outages, so that downtimes can be reduced and the availability of cloud services is improved. For this task, reinforcement learning has been proposed before~\cite{Yuan_2007_RLAAER,Ikeuchi_2020_Seq2SeqRecovery}, to explore and select remediation actions. However, having systems learn the selection of remediation actions such as migrating or restarting appliances through experimentation at runtime will not be an option in many production environments. On the other hand, reinforcement learning also might not always deal well with the large solution spaces in any case. Therefore, other works match problem cases and remediation actions based on what has successfully resolved specific issues in the past~\cite{Montani_2006_CaseBased}, even though there has not been much work on applying supervised learning methods in a similar manner, training models on actions that have successfully resolved specific problems previously. This is likely the case because training a model for a supervised approach to automatic remediation requires a sizable amount of training samples. Moreover, both case-based approaches and supervised learning methods also assume that what has worked to resolve issues in the past will work again in similar situations in the future, which is not necessarily a valid assumption for failures in large and complex cloud systems.
That is, while using ML for problem \textit{detection} has gained significant attention and produced good results, ML for problem \textit{resolution} is still only at the beginning.

Several key challenges remain for ML-supported cloud resource management and operation in general. These include the explainability and trust in model-based resource management decisions, security and safety of ML model-based cloud operations, as well as efficiently adapting large trained models to new contexts as workloads and infrastructures evolve. We, therefore, expect using monitoring data, stream processing, and ML to automate and improve the performance, dependability, and efficiency of cloud systems to continue to be a major trend in research and practice.

\begin{figure}
    \centering
    \includegraphics[width=\columnwidth]{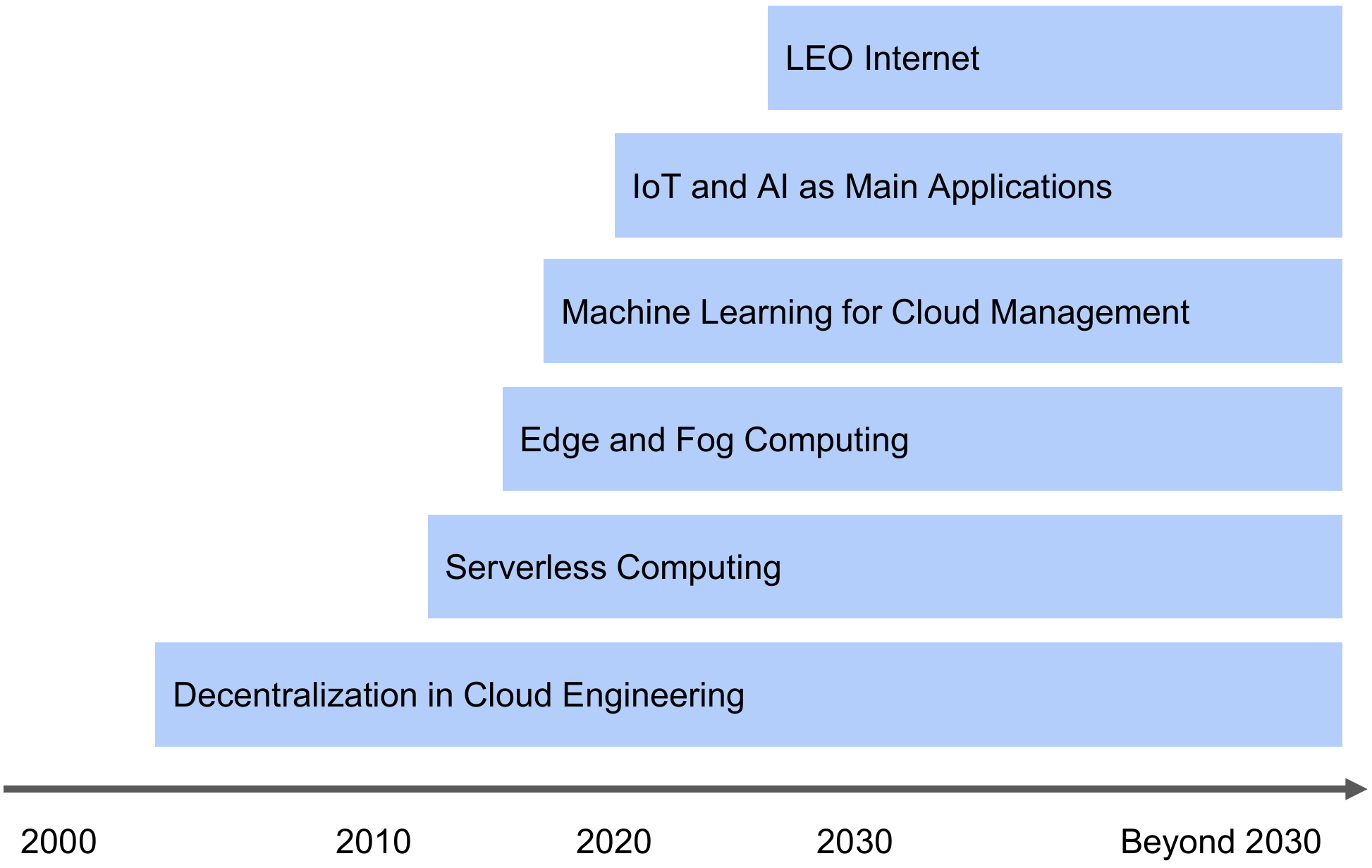}
    \caption{An attempt at a timeline of the discussed trends.}
    \label{fig:timeline}
\end{figure}

\subsection{Summary}
To recap, this paper makes the case for the following challenges and opportunities in cloud engineering over the next decade that will involve significant new research and products.

\begin{itemize}
\item Cloud as a continuum of resources from the edge to the fog to the traditional data center,
\item Constellation of Low Earth Orbit satellites providing space-based clusters of dynamically changing topology of cloud resources,
\item Increased integration of Distributed Ledgers and Blockchains with the Cloud,
\item Internet of Things and Artificial Intelligence becoming the mainstay in Cloud Computing bringing increased intelligence and automation,
\item Relieving users from deployment and provisioning challenges through increased use of serverless computing,
\item Data-driven machine learning modeling and control of critical applications
\end{itemize}
See also Figure~\ref{fig:timeline} for an overview of these trends over time.

\section{Conclusion}
This paper focuses on the importance of cloud engineering in the realm of Cloud Computing. It first lays out the contemporary landscape of cloud engineering, and then delves into the numerous challenges and opportunities for cloud engineering as new advances in both hardware and software give rise to increasingly feature-rich cloud offerings and complex distributed services.

\bibliographystyle{IEEEtran}
\bibliography{cites}

\begin{thebibliography}{100}
\providecommand{\url}[1]{#1}
\csname url@samestyle\endcsname
\providecommand{\newblock}{\relax}
\providecommand{\bibinfo}[2]{#2}
\providecommand{\BIBentrySTDinterwordspacing}{\spaceskip=0pt\relax}
\providecommand{\BIBentryALTinterwordstretchfactor}{4}
\providecommand{\BIBentryALTinterwordspacing}{\spaceskip=\fontdimen2\font plus
\BIBentryALTinterwordstretchfactor\fontdimen3\font minus
  \fontdimen4\font\relax}
\providecommand{\BIBforeignlanguage}[2]{{%
\expandafter\ifx\csname l@#1\endcsname\relax
\typeout{** WARNING: IEEEtran.bst: No hyphenation pattern has been}%
\typeout{** loaded for the language `#1'. Using the pattern for}%
\typeout{** the default language instead.}%
\else
\language=\csname l@#1\endcsname
\fi
#2}}
\providecommand{\BIBdecl}{\relax}
\BIBdecl

\bibitem{paper_lenk_cloud_landscape}
A.~Lenk, M.~Klems, J.~Nimis, S.~Tai, and T.~Sandholm, ``What's inside the
  cloud? an architectural map of the cloud landscape,'' in \emph{Proc. of ICSE
  Workshops}, 2009.

\bibitem{bermbach2011extendable}
D.~Bermbach, R.~Kern, P.~Wichmann, S.~Rath, and C.~Zirpins, ``An extendable
  toolkit for managing quality of human-based electronic services,'' in
  \emph{Proc. of HCOMP}.\hskip 1em plus 0.5em minus 0.4em\relax AAAI Press,
  2012.

\bibitem{paolacci2014inside}
G.~Paolacci and J.~Chandler, ``Inside the turk: Understanding mechanical turk
  as a participant pool,'' \emph{CDP}, 2014.

\bibitem{paper_bermbach_benchmarking_web_apis}
D.~Bermbach and E.~Wittern, ``Benchmarking web api quality,'' in \emph{Proc. of
  ICWE}.\hskip 1em plus 0.5em minus 0.4em\relax Springer, 2016.

\bibitem{paper_bermbach_webapibenchmarking2}
------, ``Benchmarking web api quality -- revisited,'' \emph{J. of Web Eng.},
  2020.

\bibitem{beck2001manifesto}
K.~Beck, M.~Beedle, A.~Van~Bennekum, A.~Cockburn, W.~Cunningham, M.~Fowler,
  J.~Grenning, J.~Highsmith, A.~Hunt, R.~Jeffries \emph{et~al.}, ``Manifesto
  for agile software development,'' 2001.

\bibitem{grambow_continuous_benchmarking_2019}
M.~Grambow, F.~Lehmann, and D.~Bermbach, ``Continuous benchmarking: Using
  system benchmarking in build pipelines,'' in \emph{Proc. of SQUEET}.\hskip
  1em plus 0.5em minus 0.4em\relax IEEE, 2019.

\bibitem{van2012kieker}
A.~Van~Hoorn, J.~Waller, and W.~Hasselbring, ``Kieker: A framework for
  application performance monitoring and dynamic software analysis,'' in
  \emph{Proc. of ICPE}, 2012.

\bibitem{paper_grambow_benchmarking_microservices}
M.~Grambow, L.~Meusel, E.~Wittern, and D.~Bermbach, ``{Benchmarking
  Microservice Performance: A Pattern-based Approach},'' in \emph{Proc. of
  SAC}.\hskip 1em plus 0.5em minus 0.4em\relax ACM, 2020.

\bibitem{grambow_benchmarking_2020}
M.~Grambow, E.~Wittern, and D.~Bermbach, ``{Benchmarking} the {Performance} of
  {Microservice} {Applications},'' \emph{SIGAPP ACR}, vol.~20, 2020.

\bibitem{roy2011efficient}
N.~Roy, A.~Dubey, and A.~Gokhale, ``{Efficient Autoscaling in the Cloud using
  Predictive Models for Workload Forecasting},'' in \emph{Proc. of
  CLOUD}.\hskip 1em plus 0.5em minus 0.4em\relax IEEE, 2011.

\bibitem{shekhar2018performance}
S.~Shekhar, H.~Abdel-Aziz, A.~Bhattacharjee, A.~Gokhale, and X.~Koutsoukos,
  ``{Performance Interference-aware Vertical Elasticity for Cloud-hosted
  Latency-sensitive Applications},'' in \emph{Proc. of CLOUD}.\hskip 1em plus
  0.5em minus 0.4em\relax IEEE, 2018.

\bibitem{bass2015devops}
L.~Bass, I.~Weber, and L.~Zhu, \emph{DevOps: A software architect's
  perspective}.\hskip 1em plus 0.5em minus 0.4em\relax Addison-Wesley, 2015.

\bibitem{kuroda2014model}
T.~Kuroda and A.~Gokhale, ``{Model-based IT Change Management for Large System
  Definitions with State-related Dependencies},'' in \emph{Proc. of
  EDOC}.\hskip 1em plus 0.5em minus 0.4em\relax IEEE, 2014.

\bibitem{bhattacharjee2018model}
A.~Bhattacharjee, Y.~Barve, A.~Gokhale, and T.~Kuroda, ``{A Model-driven
  Approach to Automate the Deployment and Management of Cloud Services},'' in
  \emph{Proc. of UCC Workshops}.\hskip 1em plus 0.5em minus 0.4em\relax IEEE,
  2018.

\bibitem{paper_kurze_cloud_federation}
T.~Kurze, M.~Klems, D.~Bermbach, A.~Lenk, S.~Tai, and M.~Kunze, ``Cloud
  federation,'' in \emph{Proc. of CLOUD COMPUTING}, 2011.

\bibitem{Salama_2014_QoSFederation}
M.~Salama and A.~Shawish, ``A qos-oriented inter-cloud federation framework,''
  in \emph{Proc. of COMPSAC}.\hskip 1em plus 0.5em minus 0.4em\relax IEEE,
  2014.

\bibitem{Rebai_2015_Federation}
S.~Rebai, M.~Hadji, and D.~Zeghlache, ``Improving profit through cloud
  federation,'' in \emph{Proc. of CCNC}.\hskip 1em plus 0.5em minus 0.4em\relax
  IEEE, 2015.

\bibitem{Hiller_2018_CPPLFederation}
J.~Hiller, M.~Kimmerlin, M.~Plauth, S.~Heikkila, S.~Klauck, V.~Lindfors,
  F.~Eberhardt, D.~Bursztynowski, J.~L. Santos, O.~Hohlfeld, and K.~Wehrle,
  ``Giving customers control over their data: Integrating a policy language
  into the cloud,'' in \emph{Proc. of IC2E}.\hskip 1em plus 0.5em minus
  0.4em\relax IEEE, 2018.

\bibitem{paper_bermbach_fog_vision}
D.~Bermbach, F.~Pallas, D.~G. Perez, P.~Plebani, M.~Anderson, R.~Kat, and
  S.~Tai, ``A research perspective on fog computing,'' in \emph{Proc. of
  ISYCC}.\hskip 1em plus 0.5em minus 0.4em\relax Springer, 2017.

\bibitem{paper_zhang_gdp}
B.~Zhang, N.~Mor, J.~Kolb, D.~S. Chan, K.~Lutz, E.~Allman, J.~Wawrzynek,
  E.~Lee, and J.~Kubiatowicz, ``The cloud is not enough: Saving iot from the
  cloud,'' in \emph{Proc. of {USENIX} {HotCloud}}, 2015.

\bibitem{paper_pfandzelter_functions_streams}
T.~Pfandzelter and D.~Bermbach, ``{IoT Data Processing in the Fog: Functions,
  Streams, or Batch Processing?}'' in \emph{Proc. of DaMove}.\hskip 1em plus
  0.5em minus 0.4em\relax IEEE, 2019.

\bibitem{paper_pallas_fog4privacy}
F.~Pallas, P.~Raschke, and D.~Bermbach, ``Fog computing as privacy enabler,''
  in \emph{Internet Computing}.\hskip 1em plus 0.5em minus 0.4em\relax IEEE,
  2020.

\bibitem{paper_pfandzelter_zero2fog}
T.~Pfandzelter, J.~Hasenburg, and D.~Bermbach, ``From zero to fog: Efficient
  engineering of fog-based internet of things applications,'' \emph{Software:
  Practice and Experience}, 2021.

\bibitem{shekhar2017indices}
S.~Shekhar, A.~D. Chhokra, A.~Bhattacharjee, G.~Aupy, and A.~Gokhale,
  ``{INDICES: Exploiting Edge Resources for Performance-aware Cloud-hosted
  Services},'' in \emph{Proc. of ICFEC}.\hskip 1em plus 0.5em minus 0.4em\relax
  IEEE, 2017.

\bibitem{shekhar2019urmila}
S.~Shekhar, A.~Chhokra, H.~Sun, A.~Gokhale, A.~Dubey, and X.~Koutsoukos,
  ``{URMILA: A Performance and Mobility-aware Fog/Edge Resource Management
  Middleware},'' in \emph{Proc. of ISORC}.\hskip 1em plus 0.5em minus
  0.4em\relax IEEE, 2019.

\bibitem{paper_george_nanolambda}
G.~George, F.~Bakir, R.~Wolski, and C.~Krintz, ``Nanolambda: Implementing
  functions as a service at all resource scales for the internet of things.''
  in \emph{Proc. of SEC}.\hskip 1em plus 0.5em minus 0.4em\relax IEEE, 2020.

\bibitem{qiu2019monocular}
X.~Qiu, A.~Keerthi, T.~Kotake, and A.~Gokhale, ``{A Monocular Vision-based
  Obstacle Avoidance Android/Linux Middleware for the Visually Impaired},'' in
  \emph{Proc. of Middleware Posters}, 2019.

\bibitem{paper_hasenburg_towards_fbase}
J.~Hasenburg, M.~Grambow, and D.~Bermbach, ``{Towards A Replication Service for
  Data-Intensive Fog Applications},'' in \emph{Proc. of SAC}.\hskip 1em plus
  0.5em minus 0.4em\relax ACM, 2020.

\bibitem{techreport_hasenburg_2019}
------, ``{FBase: A Replication Service for Data-Intensive Fog Applications},''
  in \emph{Tech. Rep. MCC.2019.1}.\hskip 1em plus 0.5em minus 0.4em\relax {TU
  Berlin, MCC Group}, 2019.

\bibitem{confais2017object}
B.~Confais, A.~Lebre, and B.~Parrein, ``An object store service for a fog/edge
  computing infrastructure based on ipfs and a scale-out nas,'' in \emph{Proc.
  of ICFEC}.\hskip 1em plus 0.5em minus 0.4em\relax IEEE, 2017.

\bibitem{paper_bermbach_cloudfederation}
D.~Bermbach, T.~Kurze, and S.~Tai, ``Cloud federation: Effects of federated
  compute resources on quality of service and cost,'' in \emph{Proc. of
  IC2E}.\hskip 1em plus 0.5em minus 0.4em\relax IEEE, 2013.

\bibitem{Pultarova2015-ml}
T.~Pultarova, ``Telecommunications - space tycoons go head to head over mega
  satell. netw. [news briefing],'' \emph{Engineering Technology}, 2015.

\bibitem{otte2013f6com}
W.~R. Otte, A.~Dubey, S.~Pradhan, P.~Patil, A.~Gokhale, G.~Karsai, and
  J.~Willemsen, ``{F6com: A component model for resource-constrained and
  dynamic space-based computing environments},'' in \emph{Proc. of
  ISORC}.\hskip 1em plus 0.5em minus 0.4em\relax IEEE, 2013.

\bibitem{levendovszky2013distributed}
T.~Levendovszky, A.~Dubey, W.~R. Otte, D.~Balasubramanian, A.~Coglio, S.~Nyako,
  W.~Emfinger, P.~Kumar, A.~Gokhale, and G.~Karsai, ``{Distributed real-time
  managed systems: A model-driven distributed secure information architecture
  platform for managed embedded systems},'' \emph{IEEE software}, 2013.

\bibitem{balasubramanian2015drems}
D.~Balasubramanian, A.~Dubey, W.~Otte, T.~Levendovszky, A.~Gokhale, P.~Kumar,
  W.~Emfinger, and G.~Karsai, ``{DREMS ML: A Wide Spectrum Architecture Design
  Language for Distributed Computing Platforms},'' \emph{Science of Computer
  Programming}, 2015.

\bibitem{Bhattacherjee2020_kr}
D.~Bhattacherjee, S.~Kassing, M.~Licciardello, and A.~Singla, ``In-orbit
  computing: An outlandish thought experiment?'' in \emph{Proc. of {ACM}
  {HotNets}}, 2020.

\bibitem{Bhosale2020_aa}
V.~Bhosale, K.~Bhardwaj, and A.~Gavrilovska, ``Toward loosely coupled
  orchestration for the {LEO} satell. edge,'' in \emph{Proc. of {USENIX}
  {HotEdge}}, 2020.

\bibitem{paper_pfandzelter_LEO_serverless}
T.~Pfandzelter, J.~Hasenburg, and D.~Bermbach, ``{Towards a Computing Platform
  for the LEO Edge},'' in \emph{Proc. of EdgeSys}.\hskip 1em plus 0.5em minus
  0.4em\relax ACM, 2021.

\bibitem{paper_pfandzelter_LEO_CDN}
T.~Pfandzelter and D.~Bermbach, ``{Edge (of the Earth) Replication: Optimizing
  Content Delivery in Large LEO Satellite Communication Networks},'' in
  \emph{Proc. of CCGrid}.\hskip 1em plus 0.5em minus 0.4em\relax IEEE, 2021.

\bibitem{web_timesmachine}
D.~Gottfrid, ``The new york times archives + amazon web services =
  timesmachine,'' in
  \emph{\url{https://open.blogs.nytimes.com/2008/05/21/the-new-york-times-archives-amazon-web-services-timesmachine/}
  (accessed on June 4, 2021)}, 2008.

\bibitem{cit-edgeMLGoogle}
\emph{Machine learning with model filtering and model mixing for edge devices
  in a heterogeneous environment},
  \url{https://patents.google.com/patent/US20160217387A1/en}, google patent,
  2016.

\bibitem{8941259}
I.~Lujic, V.~D. Maio, and I.~Brandic, ``Resilient edge data management
  framework,'' \emph{{IEEE} TSC}, 2020.

\bibitem{paper_hasenburg_disgb}
J.~Hasenburg and D.~Bermbach, ``{DisGB}: Using geo-context information for
  efficient routing in geo-distributed pub/sub systems,'' in \emph{Proc. of
  UCC}.\hskip 1em plus 0.5em minus 0.4em\relax {IEEE}, 2020.

\bibitem{paper_hasenburg_geobroker}
------, ``{GeoBroker}: Leveraging geo-context for {IoT} data distribution,''
  \emph{Elsevier Comp. Comm.}, 2020.

\bibitem{paper_hasenburg_broadcast_groups}
J.~Hasenburg, F.~Stanek, F.~Tschorsch, and D.~Bermbach, ``Managing latency and
  excess data dissemination in fog-based publish/subscribe systems,'' in
  \emph{Proc. of ICFC}.\hskip 1em plus 0.5em minus 0.4em\relax IEEE, 2020.

\bibitem{pu_low_2015}
Q.~Pu, G.~Ananthanarayanan, P.~Bodik, S.~Kandula, A.~Akella, P.~Bahl, and
  I.~Stoica, ``Low latency geo-distributed data analytics,'' in \emph{Proc. of
  SIGCOMM}.\hskip 1em plus 0.5em minus 0.4em\relax ACM, 2015.

\bibitem{heintz16mr}
B.~Heintz, A.~Chandra, R.~K. Sitaraman, and J.~Weissman, ``End-to-end
  optimization for geo-distributed mapreduce,'' \emph{IEEE TCC}, 2016.

\bibitem{vulimiri2015wanalytics}
A.~Vulimiri, C.~Curino, P.~B. Godfrey, T.~Jungblut, K.~Karanasos, J.~Padhye,
  and G.~Varghese, ``Wanalytics: Geo-distributed analytics for a data intensive
  world,'' in \emph{Proc. of SIGMOD}.\hskip 1em plus 0.5em minus 0.4em\relax
  ACM, 2015.

\bibitem{viswanathanAA16}
R.~Viswanathan, G.~Ananthanarayanan, and A.~Akella, ``Clarinet: Wan-aware
  optimization for analytics queries,'' in \emph{Proc. of OSDI}.\hskip 1em plus
  0.5em minus 0.4em\relax USENIX, 2016.

\bibitem{hung_wide_area_2018}
C.-C. Hung, G.~Ananthanarayanan, L.~Golubchik, M.~Yu, and M.~Zhang, ``Wide-area
  analytics with multiple resources,'' in \emph{Proc. of EuroSys}.\hskip 1em
  plus 0.5em minus 0.4em\relax ACM, 2018.

\bibitem{kimchi}
K.~Oh, A.~Chandra, and J.~B. Weissman, ``A network cost-aware geo-distributed
  data analytics system,'' in \emph{Proc. of {CCGRID}}.\hskip 1em plus 0.5em
  minus 0.4em\relax {IEEE}, 2020.

\bibitem{jetstream2014}
A.~Rabkin, M.~Arye, S.~Sen, V.~S. Pai, and M.~J. Freedman, ``Aggregation and
  degradation in jetstream: Streaming analytics in the wide area,'' in
  \emph{Proc. of {USENIX} {NSDI}}.\hskip 1em plus 0.5em minus 0.4em\relax
  {USENIX}, 2014.

\bibitem{heintz2015optimizing}
B.~Heintz, A.~Chandra, and R.~K. Sitaraman, ``Optimizing grouped aggregation in
  geo-distributed streaming analytics,'' in \emph{Proc. of HPDC}.\hskip 1em
  plus 0.5em minus 0.4em\relax ACM, 2015.

\bibitem{zhang_awstream_2018}
B.~Zhang, X.~Jin, S.~Ratnasamy, J.~Wawrzynek, and E.~A. Lee, ``{AWStream:
  Adaptive Wide-Area Streaming Analytics},'' in \emph{Proc. of SIGCOMM}.\hskip
  1em plus 0.5em minus 0.4em\relax ACM, 2018.

\bibitem{jonathan20wasp}
A.~Jonathan, A.~Chandra, and J.~Weissman, ``Wasp: Wide-area adaptive stream
  processing,'' in \emph{Proc. of Middleware}.\hskip 1em plus 0.5em minus
  0.4em\relax ACM, 2020.

\bibitem{albert_multi_query}
------, ``Multi-query optimization in wide-area streaming analytics,'' in
  \emph{Proc. of SoCC}.\hskip 1em plus 0.5em minus 0.4em\relax ACM, 2018.

\bibitem{heintz_trading_2016}
B.~Heintz, A.~Chandra, and R.~K. Sitaraman, ``Trading timeliness and accuracy
  in geo-distributed streaming analytics,'' in \emph{Proc. of SoCC}.\hskip 1em
  plus 0.5em minus 0.4em\relax ACM, 2016.

\bibitem{heintz_optimizing_2017}
------, ``Optimizing timeliness and cost in geo-distributed streaming
  analytics,'' \emph{{IEEE} TCC}, 2020.

\bibitem{ttl_aggregation}
D.~Kumar, J.~Li, A.~Chandra, and R.~Sitaraman, ``A ttl-based approach for data
  aggregation in geo-distributed streaming analytics,'' in \emph{ACM
  SIGMETRICS}, 2019.

\bibitem{bhattacharjee2019stratum}
A.~Bhattacharjee, Y.~Barve, S.~Khare, S.~Bao, A.~Gokhale, and T.~Damiano,
  ``{Stratum: A Serverless Framework for the Lifecycle Management of Machine
  Learning-based Data Analytics Tasks},'' in \emph{Proc. of $\{$USENIX$\}$
  {OpML}}, 2019.

\bibitem{DBLP:conf/hotos/CiparHKLGGKX13}
J.~Cipar, Q.~Ho, J.~K. Kim, S.~Lee, G.~R. Ganger, G.~Gibson, K.~Keeton, and
  E.~P. Xing, ``Solving the straggler problem with bounded staleness,'' in
  \emph{{HotOS}}.\hskip 1em plus 0.5em minus 0.4em\relax {USENIX}, 2013.

\bibitem{DBLP:journals/pomacs/AralEB20}
A.~Aral, M.~Erol{-}Kantarci, and I.~Brandic, ``Staleness control for edge data
  analytics,'' \emph{ACM POMACS}, 2020.

\bibitem{bhattacharjee2020deep}
A.~Bhattacharjee, A.~D. Chhokra, H.~Sun, S.~Shekhar, A.~Gokhale, G.~Karsai, and
  A.~Dubey, ``{Deep-Edge: An Efficient Framework for Deep Learning Model Update
  on Heterogeneous Edge},'' in \emph{Proc. of ICFEC}.\hskip 1em plus 0.5em
  minus 0.4em\relax IEEE, 2020.

\bibitem{UCB_ML_slides}
J.~E. Gonzalez, ``{AI-Systems: Machine Learning Lifecycle},'' in
  \emph{\url{https://ucbrise.github.io/cs294-ai-sys-fa19/assets/lectures/lec03/03{\_}ml-lifecycle.pdf}
  (accessed on July 22, 2021)}.

\bibitem{zhao2019dynamic}
X.~Zhao, A.~An, J.~Liu, and B.~X. Chen, ``Dynamic stale synchronous parallel
  distributed training for deep learning,'' in \emph{Proc.of ICDCS}, 2019.

\bibitem{dutta2020slow}
S.~Dutta, J.~Wang, and G.~Joshi, ``Slow and stale gradients can win the race,''
  \emph{arXiv preprint arXiv:2003.10579}, 2020.

\bibitem{li2021syncswitch_icdcs}
S.~Li, O.~Mangoubi, L.~Xu, and T.~Guo, ``Sync-switch: Hybrid parameter
  synchronization for distributed deep learning,'' in \emph{Proc.of ICDCS},
  2021.

\bibitem{Aji2017-yr}
A.~F. Aji and K.~Heafield, ``{Sparse Communication for Distributed Gradient
  Descent},'' \emph{arXiv preprint arXiv:1704.05021}, 2017.

\bibitem{Wen2017-uf}
W.~Wen, C.~Xu, F.~Yan, C.~Wu, Y.~Wang, Y.~Chen, and H.~Li, ``{TernGrad: Ternary
  Gradients to Reduce Communication in Distributed Deep Learning},''
  \emph{{NeurIPS}}, 2017.

\bibitem{Alistarh2017-ss}
D.~Alistarh, D.~Grubic, J.~Li, R.~Tomioka, and M.~Vojnovic, ``{QSGD:
  Communication-Efficient SGD via Gradient Quantization and Encoding},''
  \emph{{NeurIPS}}, 2017.

\bibitem{Thorpe2021-vh}
J.~Thorpe, Y.~Qiao, J.~Eyolfson, S.~Teng, G.~Hu, Z.~Jia, J.~Wei, K.~Vora,
  R.~Netravali, M.~Kim, and {Others}, ``{Dorylus: Affordable, Scalable, and
  Accurate GNN Training with Distributed CPU Servers and Serverless Threads},''
  in \emph{{Proc.of {OSDI}}}, 2021.

\bibitem{Wang2019-uy}
M.~Wang, C.-C. Huang, and J.~Li, ``{Supporting Very Large Models using
  Automatic Dataflow Graph Partitioning},'' in \emph{{Proc.of EuroSys}}.\hskip
  1em plus 0.5em minus 0.4em\relax ACM, 2019.

\bibitem{Carreira2019-kx}
J.~Carreira, P.~Fonseca, A.~Tumanov, A.~Zhang, and R.~Katz, ``{Cirrus: a
  Serverless Framework for End-to-end ML Workflows},'' in \emph{{Proc.of
  SoCC}}.\hskip 1em plus 0.5em minus 0.4em\relax ACM, 2019.

\bibitem{Jiang2021-wa}
J.~Jiang, S.~Gan, Y.~Liu, F.~Wang, G.~Alonso, A.~Klimovic, A.~Singla, W.~Wu,
  and C.~Zhang, ``{Towards Demystifying Serverless Machine Learning
  Training},'' in \emph{{Proc.of SIGMOD}}.\hskip 1em plus 0.5em minus
  0.4em\relax ACM, 2021.

\bibitem{Li2020-bk}
S.~Li, R.~J. Walls, and T.~Guo, ``{Characterizing and Modeling Distributed
  Training with Transient Cloud GPU Servers},'' in \emph{{Proc.of ICDCS}},
  2020.

\bibitem{Harlap2017-wa}
A.~Harlap, A.~Tumanov, A.~Chung, G.~R. Ganger, and P.~B. Gibbons, ``{Proteus:
  Agile ML Elasticity Through Tiered Reliability in Dynamic Resource
  Markets},'' in \emph{{Proc.of EuroSys}}.\hskip 1em plus 0.5em minus
  0.4em\relax ACM, 2017.

\bibitem{Peng2018-mw}
Y.~Peng, Y.~Bao, Y.~Chen, C.~Wu, and C.~Guo, ``{Optimus: An Efficient Dynamic
  Resource Scheduler for Deep Learning Clusters},'' in \emph{{Proc.of
  EuroSys}}.\hskip 1em plus 0.5em minus 0.4em\relax ACM, 2018.

\bibitem{Gu2019-hc}
J.~Gu, M.~Chowdhury, K.~G. Shin, Y.~Zhu, M.~Jeon, J.~Qian, H.~Liu, and C.~Guo,
  ``{Tiresias: A GPU Cluster Manager for Distributed Deep Learning},'' in
  \emph{{Proc.of USENIX NSDI 19}}, 2019.

\bibitem{Qiao2021-nk}
A.~Qiao, S.~K. Choe, S.~J. Subramanya, W.~Neiswanger, Q.~Ho, H.~Zhang, G.~R.
  Ganger, and E.~P. Xing, ``{Pollux: Co-adaptive Cluster Scheduling for
  Goodput-Optimized Deep Learning},'' in \emph{{Proc.of USENIX OSDI}}, 2021.

\bibitem{dlion}
R.~Hong and A.~Chandra, ``Dlion: Decentralized distributed deep learning in
  micro-clouds,'' in \emph{Proc. of HPDC}.\hskip 1em plus 0.5em minus
  0.4em\relax ACM, 2021.

\bibitem{crankshaw2020-ut}
D.~Crankshaw, G.-E. Sela, X.~Mo, C.~Zumar, I.~Stoica, J.~Gonzalez, and
  A.~Tumanov, ``{InferLine: latency-aware provisioning and scaling for
  prediction serving pipelines},'' in \emph{{Proc.of SoCC}}.\hskip 1em plus
  0.5em minus 0.4em\relax ACM, 2020.

\bibitem{Hauswald2015a}
J.~Hauswald, M.~A. Laurenzano, Y.~Zhang, C.~Li, A.~Rovinski, A.~Khurana, R.~G.
  Dreslinski, T.~Mudge, V.~Petrucci, L.~Tang, and J.~Mars, ``Sirius: An open
  end-to-end voice and vision personal assistant and its implications for
  future warehouse scale computers,'' in \emph{Proc.of ASPLOS}.\hskip 1em plus
  0.5em minus 0.4em\relax ACM, 2015.

\bibitem{Capes2017a}
T.~Capes, P.~Coles, A.~Conkie, L.~Golipour, A.~Hadjitarkhani, Q.~Hu,
  N.~Huddleston, M.~Hunt, J.~Li, M.~Neeracher \emph{et~al.}, ``Siri on-device
  deep learning-guided unit selection text-to-speech system.'' in \emph{Proc.of
  INTERSPEECH}, 2017.

\bibitem{google:translate:pipeline}
O.~Good, ``How google translate squeezes deep learning onto a phone,'' in
  \emph{Google AI Blog,
  \url{https://ai.googleblog.com/2015/07/how-google-translate-squeezes-deep.html}},
  2015.

\bibitem{bhattacharjee2019barista}
A.~Bhattacharjee, A.~D. Chhokra, Z.~Kang, H.~Sun, A.~Gokhale, and G.~Karsai,
  ``{Barista: Efficient and Scalable Serverless Serving System for Deep
  Learning Prediction Services},'' in \emph{Proc. of IC2E}.\hskip 1em plus
  0.5em minus 0.4em\relax IEEE, 2019.

\bibitem{Gilman2019-lw}
G.~R. Gilman, S.~S. Ogden, R.~J. Walls, and T.~Guo, ``{Challenges and
  Opportunities of DNN Model Execution Caching},'' in \emph{{Proc. of
  DIDL}}.\hskip 1em plus 0.5em minus 0.4em\relax ACM, 2019.

\bibitem{Romero2021-sz}
F.~Romero, Q.~Li, N.~J. Yadwadkar, and C.~Kozyrakis, ``{INFaaS: Automated
  Model-less Inference Serving},'' in \emph{{Proc.of USENIX ATC}}, 2021.

\bibitem{Fuerst2021-bz}
A.~Fuerst and P.~Sharma, ``{FaasCache: keeping serverless computing alive with
  greedy-dual caching},'' in \emph{{Proc. of ASPLOS}}.\hskip 1em plus 0.5em
  minus 0.4em\relax ACM, 2021.

\bibitem{zhou2020cost}
X.~Zhou, R.~Canady, S.~Bao, and A.~Gokhale, ``{Cost-effective Hardware
  Accelerator Recommendation for Edge Computing},'' in \emph{Proc. of
  $\{$USENIX$\}$ HotEdge}, 2020.

\bibitem{barve2019fecbench}
Y.~D. Barve, S.~Shekhar, A.~Chhokra, S.~Khare, A.~Bhattacharjee, Z.~Kang,
  H.~Sun, and A.~Gokhale, ``{FECBench: A Holistic Interference-aware Approach
  for Application Performance Modeling},'' in \emph{Proc. of IC2E}.\hskip 1em
  plus 0.5em minus 0.4em\relax IEEE, 2019.

\bibitem{Lu2020-qy}
J.~Lu, A.~Liu, F.~Dong, F.~Gu, J.~Gama, and G.~Zhang, ``{Learning under Concept
  Drift: A Review},'' \emph{IEEE Transactions on Knowledge and Data
  Engineering}, 2019.

\bibitem{Diethe2019-ag}
T.~Diethe, T.~Borchert, E.~Thereska, B.~Balle, and N.~Lawrence, ``{Continual
  Learning in Practice},'' 2019.

\bibitem{cvpr2021:WAD:talks}
``Andrej karpathy's keynote,'' \url{https://youtu.be/g6bOwQdCJrc}, 2021.

\bibitem{De_Lange2019-gp}
M.~De~Lange, R.~Aljundi, M.~Masana, S.~Parisot, X.~Jia, A.~Leonardis,
  G.~Slabaugh, and T.~Tuytelaars, ``{A continual learning survey: Defying
  forgetting in classification tasks},'' 2019.

\bibitem{paper_baldini_openwhisk}
I.~Baldini, P.~Cheng, S.~J. Fink, N.~Mitchell, V.~Muthusamy, R.~Rabbah,
  P.~Suter, and O.~Tardieu, ``{The Serverless Trilemma: Function Composition
  for Serverless Computing},'' in \emph{{Proc. of Onward!}}\hskip 1em plus
  0.5em minus 0.4em\relax ACM, 2017.

\bibitem{paper_agache_firecracker_vms}
A.~Agache, M.~Brooker, A.~Iordache, A.~Liguori, R.~Neugebauer, P.~Piwonka, and
  D.-M. Popa, ``Firecracker: Lightweight virtualization for serverless
  applications,'' in \emph{Proc. of {\{USENIX\}} {\{NSDI\}}}.\hskip 1em plus
  0.5em minus 0.4em\relax USENIX, 2020.

\bibitem{paper_pfandzelter_tinyfaas}
T.~Pfandzelter and D.~Bermbach, ``{tinyFaaS: A Lightweight FaaS Platform for
  Edge Environments},'' in \emph{{Proc. of ICFC}}.\hskip 1em plus 0.5em minus
  0.4em\relax IEEE, 2020.

\bibitem{aws-lambda}
``{AWS Lambda},'' \url{https://aws.amazon.com/lambda/}, [accessed 8-Apr-2021].

\bibitem{aws-lambda-pricing}
``{AWS Lambda Pricing Model},'' \url{https://aws.amazon.com/lambda/pricing/},
  [accessed 8-Apr-2021].

\bibitem{paper_akkus_sand}
I.~E. Akkus, R.~Chen, I.~Rimac, M.~Stein, K.~Satzke, A.~Beck, P.~Aditya, and
  V.~Hilt, ``{SAND}: Towards high-performance serverless computing,'' in
  \emph{Proc. of USENIX ATC}, 2018.

\bibitem{paper_hendrickson_openlambda}
S.~Hendrickson, S.~Sturdevant, T.~Harter, V.~Venkataramani, A.~C.
  Arpaci-Dusseau, and R.~H. Arpaci-Dusseau, ``Serverless computation with
  openlambda,'' in \emph{Proc. of USENIX HotCloud}, 2016.

\bibitem{serverless_framework}
``{Serverless Framework},'' [accessed 10-Feb-2021] \url{www.serverless.com}.

\bibitem{nsf-cps15}
``{Exploiting Parallelism and Scalability: Report on an NSF-Sponsored
  Workshop},''
  \url{http://people.duke.edu/~bcl15/documents/xps2015-report.pdf}, [accessed
  14-September-2017].

\bibitem{serverless-iot17}
G.~McGrath and P.~R. Brenner, ``{Serverless Computing: Design, Implementation,
  and Performance},'' in \emph{{Proc. of ICDCS Workshops}}, 2017.

\bibitem{paper_manner_coldstarts}
J.~Manner, M.~Endress, T.~Heckel, and G.~Wirtz, ``Cold start influencing
  factors in function as a service,'' in \emph{Proc. of {UCC} Workshops}.\hskip
  1em plus 0.5em minus 0.4em\relax IEEE, 2018.

\bibitem{paper_puripunpinyo_java}
H.~Puripunpinyo and M.~H. Samadzadeh, ``Effect of optimizing java deployment
  artifacts on aws lambda,'' in \emph{Proc. of DCPerf}.\hskip 1em plus 0.5em
  minus 0.4em\relax IEEE, 2017.

\bibitem{paper_lloyd_microservice}
W.~Lloyd, S.~Ramesh, S.~Chinthalapati, L.~Ly, and S.~Pallickara, ``Serverless
  computing: An investigation of factors influencing microservice
  performance,'' in \emph{Proc. of IC2E}.\hskip 1em plus 0.5em minus
  0.4em\relax IEEE, 2018.

\bibitem{paper_lloyd_serverless}
W.~Lloyd, M.~Vu, B.~Zhang, O.~David, and G.~Leavesley, ``Improving application
  migration to serverless computing platforms: Latency mitigation with
  keep-alive workloads,'' in \emph{Proc. of WoSC}.\hskip 1em plus 0.5em minus
  0.4em\relax IEEE, 2018.

\bibitem{paper_hellerstein_serverless}
J.~M. Hellerstein, J.~Faleiro, J.~E. Gonzalez, J.~Schleier-Smith, V.~Sreekanti,
  A.~Tumanov, and C.~Wu, ``Serverless computing: One step forward, two steps
  back,'' \emph{Proc. of CIDR}, 2019.

\bibitem{paper_sreekanti_cloudburst}
V.~Sreekanti, C.~W. X.~C. Lin, J.~M. Faleiro, J.~E. Gonzalez, J.~M.
  Hellerstein, and A.~Tumanov, ``Cloudburst: Stateful functions-as-a-service,''
  \emph{Proc. of VLDB}, 2020.

\bibitem{paper_bermbach_auctions4function_placement}
D.~Bermbach, S.~Maghsudi, J.~Hasenburg, and T.~Pfandzelter, ``Towards
  auction-based function placement in serverless fog platforms,'' in
  \emph{Proc. of ICFC}.\hskip 1em plus 0.5em minus 0.4em\relax IEEE, 2020.

\bibitem{paper_baresi_serverless_fog}
L.~Baresi and D.~F. Mendonca, ``Towards a serverless platform for edge
  computing,'' in \emph{Proc. of ICFC}.\hskip 1em plus 0.5em minus 0.4em\relax
  IEEE, 2019.

\bibitem{paper_jain_splitserve}
A.~Jain, A.~F. Baarzi, G.~Kesidis, B.~Urgaonkar, N.~Alfares, and M.~Kandemir,
  ``Splitserve: Efficiently splitting apache spark jobs across faas and iaas,''
  in \emph{Proc. of Middleware}, 2020.

\bibitem{paper_bermbach_faas_coldstarts}
D.~Bermbach, A.-S. Karakaya, and S.~Buchholz, ``{Using Application Knowledge to
  Reduce Cold Starts in FaaS Services},'' in \emph{Proc. of SAC}.\hskip 1em
  plus 0.5em minus 0.4em\relax ACM, 2020.

\bibitem{paper_ristov_afcl}
S.~Ristov, S.~Pedratscher, and T.~Fahringer, ``Afcl: An abstract function
  choreography language for serverless workflow specification,'' \emph{Elsevier
  FGCS}, 2021.

\bibitem{paper_scheuner_faas_benchmarking}
J.~Scheuner and P.~Leitner, ``Function-as-a-service performance evaluation: A
  multivocal literature review,'' \emph{Elsevier JSS}, 2020.

\bibitem{paper_grambow_befaas}
M.~Grambow, T.~Pfandzelter, L.~Burchard, C.~Schubert, M.~Zhao, and D.~Bermbach,
  ``{BeFaaS: An Application-Centric Benchmarking Framework for FaaS
  Platforms},'' in \emph{Proc. of IC2E}.\hskip 1em plus 0.5em minus 0.4em\relax
  IEEE, 2021.

\bibitem{web_developers_worldwide}
{Daxx}, ``{How Many Software Developers Are There in the World?}'' in
  \emph{https://www.daxx.com/blog/development-trends/number-software-developers-world},
  2020.

\bibitem{armbrust2010-vc}
M.~Armbrust, A.~Fox, R.~Griffith, A.~D. Joseph, R.~Katz, A.~Konwinski, G.~Lee,
  D.~Patterson, A.~Rabkin, I.~Stoica, and M.~Zaharia, ``A view of cloud
  computing,'' \emph{CACM}, 2010.

\bibitem{web_asgard_spinnaker}
{Netflix Tech Blog}, ``Moving from asgard to spinnaker,'' in
  \emph{https://netflixtechblog.com/moving-from-asgard-to-spinnaker-a000b2f7ed17},
  2015.

\bibitem{web_titus}
------, ``The evolution of container usage at netflix,'' in
  \emph{https://netflixtechblog.com/the-evolution-of-container-usage-at-netflix-3abfc096781b},
  2017.

\bibitem{OccupyCloud2017}
E.~Jonas, Q.~Pu, S.~Venkataraman, I.~Stoica, and B.~Recht, ``Occupy the cloud:
  distributed computing for the 99{\%},'' in \emph{Proc. of SoCC}.\hskip 1em
  plus 0.5em minus 0.4em\relax {ACM}, 2017.

\bibitem{Venkataraman_Ernest_2016}
S.~Venkataraman, Z.~Yang, M.~Franklin, B.~Recht, and I.~Stoica, ``{Ernest:
  Efficient Performance Prediction for Large-Scale Advanced Analytics},'' in
  \emph{{Proc. of USENIX NSDI}}.\hskip 1em plus 0.5em minus 0.4em\relax USENIX,
  2016.

\bibitem{Alipourfard_CherryPick_2017}
O.~Alipourfard, H.~H. Liu, J.~Chen, S.~Venkataraman, M.~Yu, and M.~Zhang,
  ``{CherryPick: Adaptively Unearthing the Best Cloud Configurations for Big
  Data Analytics},'' in \emph{{Proc. of USENIX NSDI}}.\hskip 1em plus 0.5em
  minus 0.4em\relax USENIX, 2017.

\bibitem{Thamsen_Ellis_2017}
L.~Thamsen, I.~Verbitskiy, J.~Beilharz, T.~Renner, A.~Polze, and O.~Kao,
  ``{Ellis: Dynamically Scaling Distributed Dataflows to Meet Runtime
  Targets},'' in \emph{{Proc. of CloudCom}}.\hskip 1em plus 0.5em minus
  0.4em\relax IEEE, 2017.

\bibitem{Hongzi_2016_DeepRM}
H.~Mao, M.~Alizadeh, I.~Menache, and S.~Kandula, ``Resource management with
  deep reinforcement learning,'' in \emph{Proc.of {HotNets}}.\hskip 1em plus
  0.5em minus 0.4em\relax ACM, 2016.

\bibitem{Cheong_2019_SCARL}
M.~Cheong, H.~Lee, I.~Yeom, and H.~Woo, ``Scarl: Attentive reinforcement
  learning-based scheduling in a multi-resource heterogeneous cluster,''
  \emph{IEEE Access}, 2019.

\bibitem{Thamsen_2020_MaryHugoHugo}
L.~Thamsen, J.~Beilharz, V.~T. Tran, S.~Nedelkoski, and O.~Kao, ``Mary, hugo,
  and hugo\textasteriskcentered: Learning to schedule distributed data-parallel
  processing jobs on shared clusters,'' \emph{Concurrency and Computation:
  Practice and Experience}, 2020.

\bibitem{Qiu_2020_FIRM}
H.~Qiu, S.~S. Banerjee, S.~Jha, Z.~T. Kalbarczyk, and R.~K. Iyer, ``{FIRM}: An
  intelligent fine-grained resource management framework for slo-oriented
  microservices,'' in \emph{Proc. of {OSDI})}.\hskip 1em plus 0.5em minus
  0.4em\relax USENIX, 2020.

\bibitem{Delimitrou_Paragon_2013}
C.~Delimitrou and C.~Kozyrakis, ``{Paragon: QoS-aware Scheduling for
  Heterogeneous Datacenters},'' in \emph{{Proc. of ASPLOS}}.\hskip 1em plus
  0.5em minus 0.4em\relax ACM, 2013.

\bibitem{Delimitrou_Quasar_2014}
------, ``{Quasar: Resource-Efficient and QoS-aware Cluster Management},'' in
  \emph{{Proc. of ASPLOS}}.\hskip 1em plus 0.5em minus 0.4em\relax ACM, 2014.

\bibitem{Gulenko_2016_MLforClouds}
A.~Gulenko, M.~Wallschl{\"a}ger, F.~Schmidt, O.~Kao, and F.~Liu, ``Evaluating
  machine learning algorithms for anomaly detection in clouds,'' in
  \emph{Proc.of {Big Data}}.\hskip 1em plus 0.5em minus 0.4em\relax IEEE, 2016.

\bibitem{Huang_2017_TSA}
C.~Huang, G.~Min, Y.~Wu, Y.~Ying, K.~Pei, and Z.~Xiang, ``Time series anomaly
  detection for trustworthy services in cloud computing systems,'' \emph{IEEE
  Transactions on Big Data}, 2017.

\bibitem{Ibidunmoye_2018_AAD}
O.~Ibidunmoye, A.-R. Rezaie, and E.~Elmroth, ``Adaptive anomaly detection in
  performance metric streams,'' \emph{IEEE Transactions on Network and Service
  Management}, 2018.

\bibitem{Nedelkoski_2020_Logsy}
S.~Nedelkoski, J.~Bogatinovski, A.~Acker, J.~Cardoso, and O.~Kao,
  ``Self-attentive classification-based anomaly detection in unstructured
  logs,'' in \emph{Proc. of ICDM}.\hskip 1em plus 0.5em minus 0.4em\relax IEEE,
  2020.

\bibitem{Wu_2020_MicroRCA}
L.~Wu, J.~Tordsson, E.~Elmroth, and O.~Kao, ``Microrca: Root cause localization
  of performance issues in microservices,'' in \emph{Proc. of NOMS}.\hskip 1em
  plus 0.5em minus 0.4em\relax IEEE, 2020.

\bibitem{Scheinert_2020_Telesto}
D.~Scheinert and A.~Acker, ``Telesto: A graph neural network model for anomaly
  classification in cloud services,'' in \emph{Proc. of ICSOC}.\hskip 1em plus
  0.5em minus 0.4em\relax Springer, 2020.

\bibitem{Yuan_2007_RLAAER}
C.~Yuan and Q.~Zhu, ``A reinforcement learning approach to automatic error
  recovery,'' in \emph{Proc.of DSN}.\hskip 1em plus 0.5em minus 0.4em\relax
  IEEE, 2007.

\bibitem{Ikeuchi_2020_Seq2SeqRecovery}
H.~Ikeuchi, A.~Watanabe, T.~Hirao, M.~Morishita, M.~Nishino, Y.~Matsuo, and
  K.~Watanabe, ``Recovery command generation towards automatic recovery in ict
  systems by seq2seq learning,'' in \emph{Proc.of NOMS}.\hskip 1em plus 0.5em
  minus 0.4em\relax IEEE, 2020.

\bibitem{Montani_2006_CaseBased}
S.~Montani and C.~Anglano, ``Case-based reasoning for autonomous service
  failure diagnosis and remediation in software systems,'' in \emph{Proc. of
  ECCBR}.\hskip 1em plus 0.5em minus 0.4em\relax Springer, 2006.

\end{thebibliography}

\end{document}